\documentclass[10pt,a4paper]{article}

\PassOptionsToPackage{cmex10}{amsmath}

\usepackage[left=2.8cm,right=2.8cm,top=3.0cm,bottom=3.0cm]{geometry}

\usepackage{hyphenat}
\usepackage{enumerate}
\usepackage{rotating}
\usepackage{multirow}
\usepackage[T1]{fontenc}
\usepackage{enumitem}
\usepackage[utf8]{inputenc}
\usepackage{subcaption}
\usepackage{todonotes}
\usepackage{tikz}
\usepackage{adjustbox}
\usepackage{algorithm}
\usepackage[noend]{algpseudocode}
\usepackage{caption}
\usepackage{colortbl}
\usepackage[cmex10]{amsmath}
\usepackage{amsfonts}
\usepackage{array}
\usepackage{authblk}
\usepackage[hyphens]{url}

\algdef{SE}[SUBALG]{Indent}{EndIndent}{}{\algorithmicend\ }%
\algtext*{Indent}
\algtext*{EndIndent}

\RequirePackage[]{acro}
\NewDocumentCommand\acrodef{mO{#1}mG{}}{\DeclareAcronym{#1}{short={#2}, long={#3}, #4}}
\acrodef{ABS}{Agent-Based Simulation}
\acrodef{ALU}{Arithmetic Logic Units}
\acrodef{APU}{Accelerated Processing Unit}
\acrodef{ASIC}{Application Specific Integrated Circuit}
\acrodef{CLB}{Configurable Logic Block}
\acrodef{CPU}{Central Processing Unit}
\acrodef{CUDA}{Compute Unified Device Architecture}
\acrodef{DES}{Discrete Event Simulation}
\acrodef{DSP}{Digital Signal Processing}
\acrodef{EDA}{Electronic Design Automation}
\acrodef{FPGA}{Field-Programmable Gate Array}
\acrodef{GPC}{Graphics Processing Cluster}
\acrodef{GPU}{Graphics Processing Unit}
\acrodef{GPGPU}{General-Purpose Computing on Graphics Processing Units}
\acrodef{HDL}{Hardware Description Language}
\acrodef{LUT}{Look-Up Table}
\acrodef{SIMD}{Single Instruction, Multiple Data}
\acrodef{SFU}{Special Function Unit}
\acrodef{SM}{Streaming Multiprocessor}
\acrodef{SP}{Streaming Processor}
\acrodef{TDP}{Thermal Design Power}

\makeatletter
\newcommand{\manuallabel}[2]{\def\@currentlabel{#2}\label{#1}}
\makeatother

\date{}
\begin{document}

\title{A Survey on Agent-based Simulation using\\ Hardware Accelerators}
%

%
\author[1,3]{Jiajian Xiao}
\author[1,2]{Philipp Andelfinger}
\author[1,3]{David Eckhoff}
\author[2]{\\Wentong Cai}
\author[3]{Alois Knoll}
\affil[1]{TUMCREATE, Singapore}
\affil[2]{School of Computer Science and Engineering, Nanyang Technological University}
\affil[3]{Department of Informatics, Technische Universit\"{a}t M\"{u}nchen}

\maketitle

\begin{abstract}
Due to decelerating gains in single-core CPU performance, computationally expensive simulations are increasingly executed on highly parallel hardware platforms.
Agent-based simulations, where simulated entities act with a certain degree of autonomy, frequently provide ample opportunities for parallelisation.
Thus, a vast variety of approaches proposed in the literature demonstrated considerable performance gains using hardware platforms such as many-core CPUs and GPUs, merged CPU-GPU chips as well as FPGAs.
Typically, a combination of techniques is required to achieve high performance for a given simulation model, putting substantial burden on modellers.
To the best of our knowledge, no systematic overview of techniques for agent-based simulations on hardware accelerators has been given in the literature.
To close this gap, we provide an overview and categorization of the literature according to the applied techniques.
Since at the current state of research, challenges such as the partitioning of a model for execution on heterogeneous hardware are still a largely manual process, we sketch directions for future research towards automating the hardware mapping and execution.
This survey targets modellers seeking an overview of suitable hardware platforms and execution techniques for a specific simulation model, as well as methodology researchers interested in potential research gaps requiring further exploration.
\end{abstract}


%

\section{Introduction}
Since around 2005, it can be observed that due to the breakdown of Dennard scaling, clock frequencies of single CPUs are no longer increasing significantly, even though the transistor counts are still growing~\cite{sutter2005free}.
Instead, CPU manufacturers have more and more focused on developing multi-core processors.
This in turn calls for parallel computing techniques, as programmes (including simulations) that cannot be run in parallel can no longer simply be sped up by incorporating a newer and faster CPU.
Performance can be increased further when the workload of a programme is efficiently distributed to heterogeneous hardware such as Graphics Processing Units (GPUs) or Field Programmable Gate Arrays (FPGAs)~\cite{che2008accelerating}.

Some types of hardware are better suited for certain tasks than others, for example, tasks with large amounts of fine-grained parallelism can benefit greatly from the massively parallel architecture of modern GPUs with its thousands of cores.
Tasks that are largely sequential or characterised by unpredictable data accesses and control flow lend themselves better to CPUs with out-of-order execution, long pipelines and large caches.
Similarly, if offloading a task to a GPU requires copying large amounts of data to and from graphics memory, execution on a CPU may be preferable even if substantial parallelism is available.
This issue can be addressed by an Accelerated Processing Unit (APU), where CPU and an integrated graphics core (of lower performance compared to stand-alone GPUs) share the same memory.
Lastly, compute-intensive and memory-light tasks can be outsourced to FPGAs which can be programmed to carry out specific computations in hardware.

One field that has always sought after more performance is the field of simulation.
Faster computers allow an increase in complexity of the incorporated simulation models, allowing researchers to obtain more accurate results in a faster manner.
Agent-based simulations have received broad attention as they can be employed to study various domains, such as road traffic~\cite{doniec2008behavioral}, social networks~\cite{epstein1999agent}, pedestrian movement~\cite{teknomo2016review}, military~\cite{cioppa2004military}, biology~\cite{an2009agent}, economics~\cite{tesfatsion2006agent} and so on.
The main characteristic of agent-based simulation is that autonomous agents (e.g., individuals or entities) act and interact to create effects of emergence on the entire system.
The complex decision-making of agents and the huge scale of many simulated systems can lead to enormous runtimes, motivating the need for employing high-performance computing platforms.

Agent-based simulations are a promising target for parallel computing techniques as agents are autonomous and in some cases carry out independent computations.
In mobility simulations, interactions between agents usually only take place between close-by agents in a somewhat regular 2D or 3D environment, allowing researchers to employ space partitioning without inducing too much synchronisation overhead.
Moreover, many ABS are time-stepped and agents are often updated at the same logical time, providing inherent independence and thus potentials for parallelised execution.
Unfortunately, being able to partition a problem and execute it in parallel is not a guarantee that it can be accelerated using heterogeneous hardware.

To enable ABS on heterogeneous hardware, some general challenges have to be overcome.
First, the simulation has to be partitioned with heterogeneity in mind to decide which part of the program lends itself best to a specific hardware device, considering the resulting overhead from data transfers between the different devices.
From this it follows that depending on the used hardware, the mapping of simulation parts to hardware devices will likely be different.
Complex simulations typically also exhibit scattered and unpredictable memory access and control flow as the model state develops dynamically over time.
This further complicates an efficient distribution to heterogeneous hardware.
Lastly, in order to make heterogeneous accelerators available to modellers even without having in-depth knowledge of the specific hardware platforms, there is a need for frameworks that abstract away from hardware specifics.
Some of the common frameworks provide variants supporting parallel and distributed execution, e.g., MASON~\cite{luke2005mason}, Repast-HPC~\cite{dubitzky2011repast}, EcoLab~\cite{standish2004ecolab}, and GridABM~\cite{szemes2010gridabm}. However, these frameworks only support traditional CPU-based environments.
Some frameworks such as FLAME GPU~\cite{Coakley-2016} and MCMAS~\cite{laville2013mcmas} have been proposed that focus on the execution on specific accelerators such as graphics cards.
   
In this survey, we structure the complex landscape of agent-based simulation on heterogeneous hardware.
We give an overview of existing types of hardware that have been employed to accelerate agent-based simulations and discuss past developments and current trends.
While some surveys exist that present generic high-performance computing techniques using heterogeneous hardware~\cite{vestias2014trends,mittal2015survey,escobar2016suitability}, we highlight the specific challenges of ABS on heterogeneous hardware and categorize an ample body of related work along these challenges.
For each challenge, we discuss in detail how existing literature has contributed to solving them.
This overview allows us to identify research gaps that need to be filled in order to establish heterogeneous accelerators in the simulation domain and making them applicable to a wider range of problems -- ideally by providing an automated process to support the modeller.

The remainder of this survey is structured as follows: in Section~\ref{sec:hardware_platforms}, we characterise the main classes of hardware accelerators for general-purpose computations. Section~\ref{sec:abs} provides an overview of agent-based simulation concepts and outlines the computational challenges of executing agent-based simulations on hardware accelerators. In Section~\ref{sec:addressing_the_challenges}, we systematise and survey the existing works according to the identified challenges and according to the techniques used to do so. In Section~\ref{sec:towards_an_automated_offloading_procedure}, we discuss unresolved challenges and outline how a system tackling these challenges could look like, thus sketching avenues for future work. Section~\ref{sec:conclusions} summarizes our findings and concludes the survey.

\section{Hardware Platforms}
\label{sec:hardware_platforms}
In this section, we describe the technical characteristics, the benefits as well as the limitations of hardware platforms that have been used to accelerate agent-based simulations. 
We focus on many-core \acsp{CPU}, \acp{GPU}, \acp{APU}, and \acp{FPGA}.
Readers familiar with these hardware platforms may skip this section and continue to Section~\ref{sec:abs}.

\subsection{Many-Core \acp{CPU}}
\textbf{Architecture:}
A many-core (or many integrated core, MIC) CPU contains a group of CPU cores on a single chip.
One of the well-known many-core CPUs, the Intel Xeon Phi, is equipped with up to 72 x86-compatible CPU cores communicating via an internal Network-on-Chip which enables fast and parallel data transfer between the cores.
\begin{figure}[b]
	\centering
	\includegraphics[width=7cm]{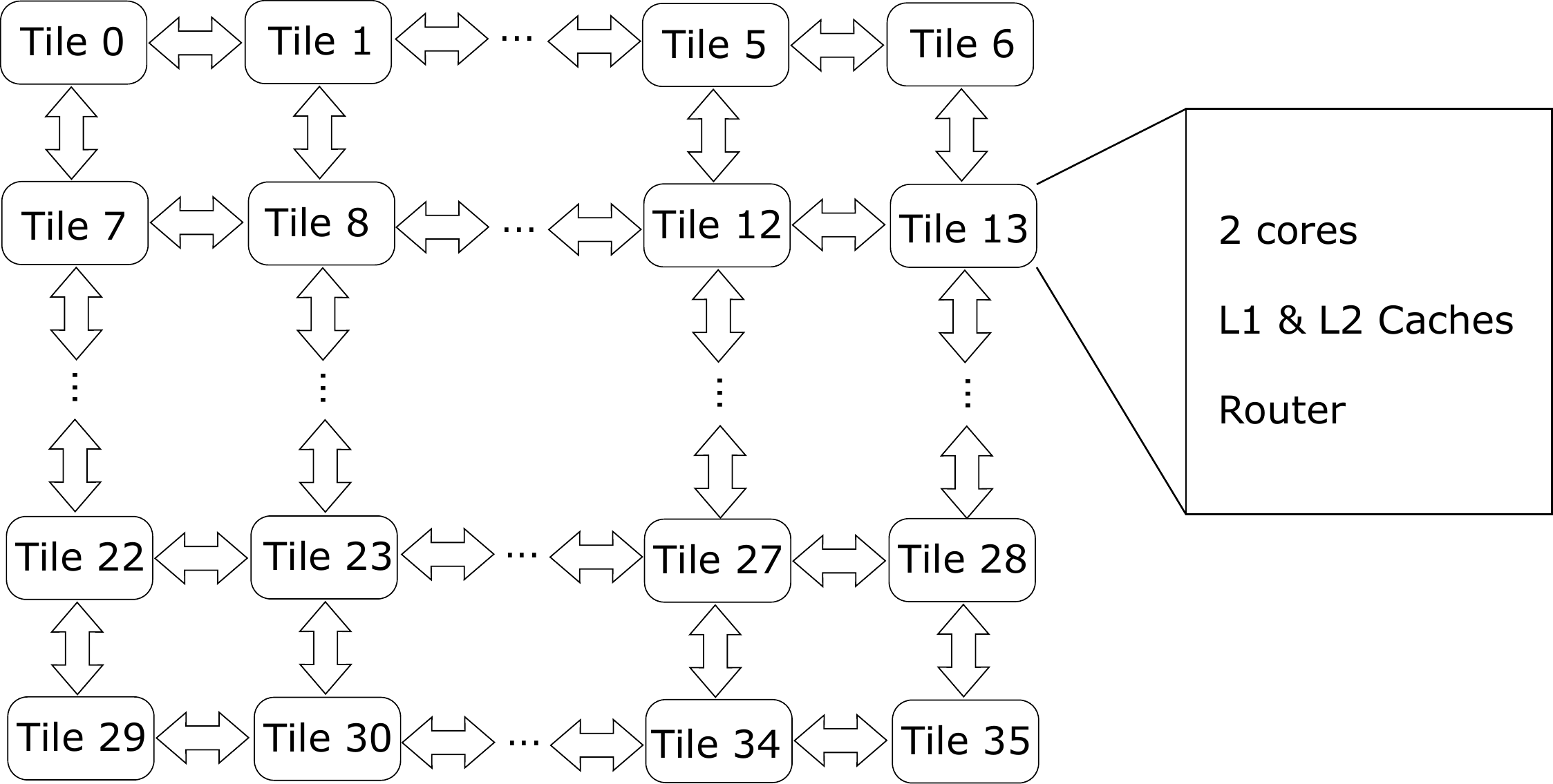}
	\caption{The tile-based architecture of the Intel Xeon Phi 7290F Processor based on Knights Landing~\cite{intelknightlanding}.}
	\label{fig:manycore}
\end{figure} 
A many-core CPU can be connected to the host machine via PCI-E or can be a standalone \ac{CPU} with direct access to the system memory.
Figure~\ref{fig:manycore} shows an overview of the second generation Intel Xeon Phi 7290F (Knights Landing) processor with its 72 cores that are grouped into 36 tiles interconnected by a 2D mesh channel.
Each tile has 2 cores sharing 1MB of L2 cache.
All L2 caches are kept fully coherent by a distributed tag directory.
The processor supports a maximum of 384GB of DDR4 RAM. In addition, 16GB of 3D-stacked multi-channel DRAM can be used for transparent caching or managed manually.

In the past years, a number of non-x86 many-core CPUs have emerged, such as the Parallella Board~\cite{agathos2015targeting}, the Epiphany-V~\cite{olofsson2016epiphany}, and the Kalray MPPA (Massively Parallel Processor Array)~\cite{de2013clustered}.

\textbf{Benefits:}
A notable advantage of some many-core CPUs over GPUs and FPGAs is their capability to execute largely unmodified code written for regular CPUs~\cite{lai2013accelerating}.
This makes migration to these platforms easier, given the code is highly parallelisable.
Since the individual cores support out-of-order execution, employ deep instruction pipelines, and have access to comparatively large caches, the need to adapt a program's control flow to the hardware is less pressing than with, e.g., GPUs~\cite{brodtkorb2013graphics}.
Still, as some many-core processors support vector operations through instruction set extensions such as AVX-512~\cite{intelavx512instructions}, a single-instruction, multiple-data (SIMD) style of programming can extract further parallelism.

Recent work showed that many-core CPUs can substantially accelerate \ac{DES}~\cite{Jagtap-2012,Williams-2017}.
A number of authors also evaluated the acceleration of various types of simulations such as fluid dynamics and seismic wave propagation using non-x86 many-cores~\cite{Raase-2015,castro2016seismic}.

\textbf{Limitations:}
In light of the comparatively high cost of recent many-core CPUs($\approx$ US\$3368.00 as of 03/2018 for an Intel Xeon Phi Processor 7290F) compared to other accelerators, the performance gains compared to traditional multi-core CPUs have frequently been relatively low.  
Even when optimising scientific code for a many-core CPU, there may only be a single-digit speedup over an execution on a traditional multi-core CPU, while in some cases there may even be an increase in runtime~\cite{Barnes-2016}.
Further, since the performance depends strongly on parameters such as the number of threads and on employing the different types of memory available on a many-core CPU, it necessary to tune these aspects to the given problem and hardware~\cite{Liu-2017a}.

\subsection{\acfp{GPU}}
\textbf{Architecture:}
\acp{GPU} utilise a massively parallel architecture, which makes them considerably more efficient than general purpose \acp{CPU} when large volumes of data can be processed in parallel.
Their original purpose was to accelerate the processing of three-dimensional scenes to be displayed on two-dimensional screens. However, modern \acp{GPU} have evolved to support a wide range of computational tasks.

The evolution of \acp{GPU} (and with that their applicability for simulation) is characterised by three essential steps.
In the 1990s, \acp{GPU} followed a fixed-function architecture, which processed a scene's geometry to produce the colour and transparency values for each of the screen's pixels in a pipelined fashion.
In 2001, Nvidia released the GeForce 3, a new GPU generation which marks the second stage of GPU evolution.
The GeForce 3 included so-called shader units, which execute programs applied to large numbers of pixel RGBA values or vertices of the objects in a 3D scene.

The flexibility of shader programming made the idea of \ac{GPGPU} practical, with early GPGPU work mapping raw data to pixels or vertices to achieve GPU-based parallel programming.
Finally, in 2006, the shader architecture was unified by no longer distinguishing between vertex and pixel shaders.
Now, with the \ac{CUDA} programming framework, it became possible for \acp{GPU} to seamlessly perform general-purpose computational tasks~\cite{ModernGPU2008,CUDA-2018}.


\begin{figure}[t]
	\centering
	\includegraphics[width=7.5cm]{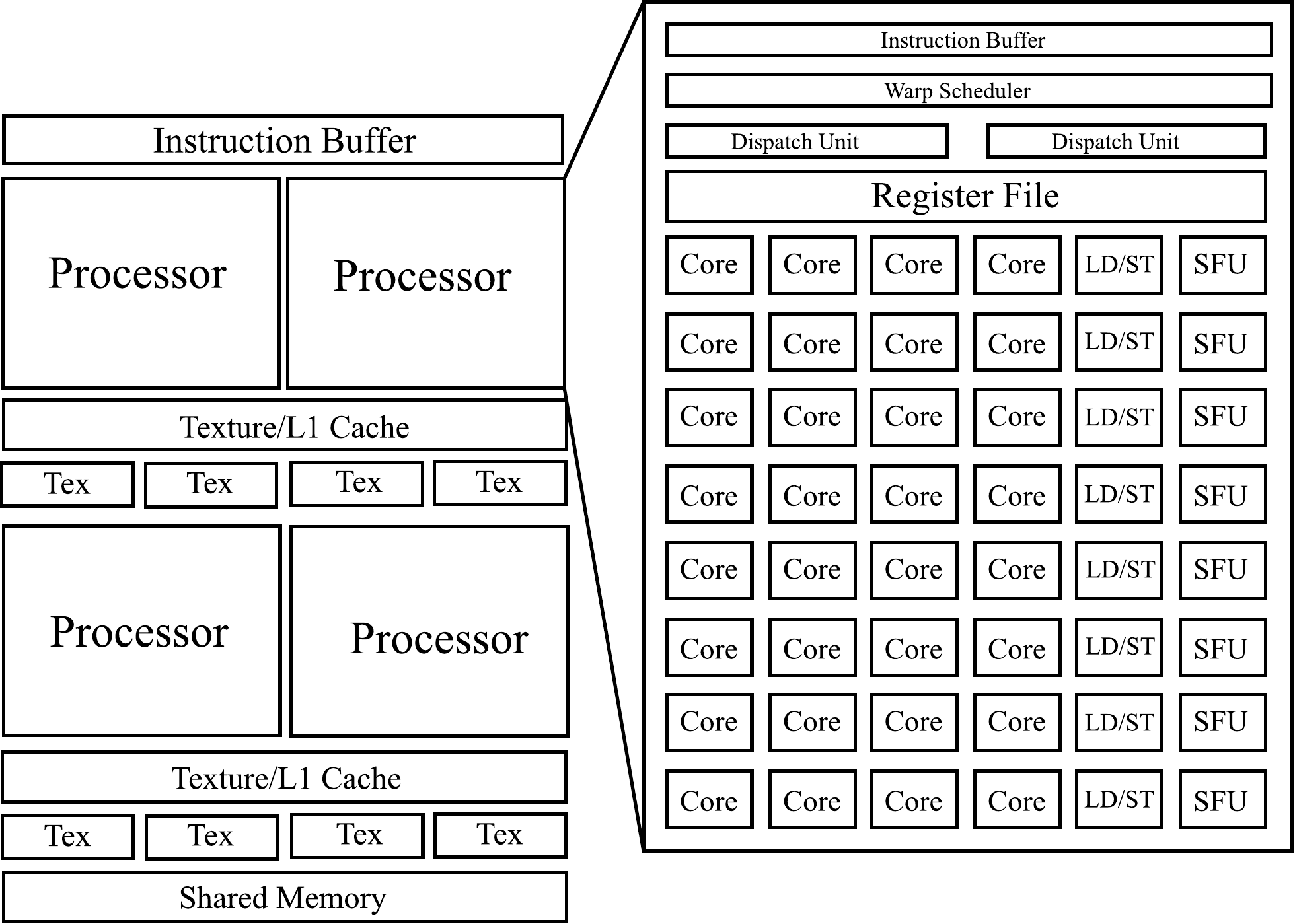}
	\caption{A Streaming Multiprocessor (SM) in a GP105 GPU based on Nvidia's Pascal architecture~\cite{GeForce_GTX_1080}.}
	\label{fig:gpusm}
\end{figure}

We sketch the GPU architecture and programming model on the basis of Nvidia's  terminology.
AMD hardware follows a similar design.
A modern GPU consists of a scalable number of \acp{SM}, which contain a number of \acp{SP} that perform most of the computations, \acp{SFU} to efficiently perform special operations such as executing trigonometric functions, on-chip memory, cache and registers.
In addition, modern \acp{GPU} have L2 cache and off-chip RAM, both of which are shared among all \acp{SM}~\cite{GeForce_GTX_1080}. Nvidia's GeForce GTX1080, for example, has 20 \acp{SM}, each of them containing 128 \acp{SP}, 32 \acp{SFU}, 256KB of registers, 8 texture units, 96KB of low-latency memory, and 48KB of L1 cache Figure~\ref{fig:gpusm} . There are 2 MB of L2 cache and 8 GB of off-chip RAM, referred to as global memory.

CUDA~\cite{CUDA-2018} (supporting Nvidia \acp{GPU}), OpenACC~\cite{openacc2011openacc} and OpenCL~\cite{Stone-2010} (the latter two supporting Nvidia, AMD \acp{GPU}, and Intel \acp{CPU} and integrated GPU) are common programming frameworks for GPGPU.
Both CUDA and OpenCL follow a similar programming model, with some differences in terminology.

The work to be performed by a GPU program is organized in a hierarchical fashion, aligned with the properties of the underlying hardware: at the lowest level, there are threads representing a sequential control flow.
On a logical level, all threads execute the same GPU program in parallel.
Threads are grouped into \emph{warps} of a hardware-specific size (32 threads on Nvidia hardware).
Within each warp, threads execute in lockstep, i.e., if the control flow among threads within a warp diverges, the different branches are serialised.
Thus, although the serialisation is transparent to the programmer, it is important to minimise intra-warp divergence.
A configurable number of warps forms a block.
Warps inside a block have access to a limited amount of low-latency shared memory and can synchronize efficiently.
Blocks are assigned to an SM persistently, i.e., the required registers and shared memory are assigned to the block until all warps have finished execution.
Per-SM warp schedulers dynamically assign runnable warps to the available \acp{SP} to minimise stalling on high-latency memory accesses.
Typically, there are many more threads than physical \acp{SP}, providing ample opportunities for this type of memory latency hiding~\cite{CUDA-2018}.

A key aspect when programming GPUs is the optimisation of memory access patterns.
The GPU hardware prescribes certain rules according to which memory accesses multiple threads can be \emph{coalesced}, i.e., executed in aggregate~\cite{CUDA-2018}.
Generally, the number of physical memory transactions required is minimised when threads with adjacent logical indexes access adjacent locations in memory.
Since many applications require scattered or even unpredictable memory access, achieving memory coalescing is a common focus of works in parallel programming on GPUs (e.g.,~\cite{wu2002calibration,fauzia2015characterizing}).

The recent Nvidia Volta architecture provides individual threads with their own execution context, enabling a more fine-grained control over the intra-warp control flow~\cite{Nvidia_volta_architecture}.

\textbf{Benefits:}
The hardware and programming model of GPUs lend itself well to problems that can be expressed so that large numbers of similar or identical operations are performed on different data.
Commonly, GPUs accelerate fine-grained data-parallel tasks by one to two orders of magnitude compared to an implementation on multi-core CPUs.

Mature GPU programming frameworks such as CUDA, OpenCL and OpenACC enable relatively simple programming compared to other accelerators such as FPGAs~\cite{falsafi2017fpgas}.
Libraries such as Thrust~\cite{bell2011thrust} and CUBLAS~\cite{nvidia2012cublas} supply the programmer with high-performance implementations for common tasks such as parallel reduction, sorting, and linear algebra operations.
Programming frameworks are available even for more specialised tasks such as agent-based simulation~\cite{Coakley-2016}.

Beside the performance benefits of GPUs, Richmond and Romano~\cite{Richmond-2008} emphasise the opportunities for efficient visualisation of simulations. Since the agent data is already stored in graphics memory, visualisation can be achieved easily by passing the agent data to vertex or texture buffers.

\textbf{Limitations:}
Most current GPUs are connected to their host CPU via the PCI-E bus. Thus, the GPU does not have direct access to the host memory.
Data transfer between CPU and GPU is expensive in terms of latency and should therefore be reduced as much as possible.
For instance, according to its specification, a PCI-E 3.0 x16 connection allows an Nvidia Titan X card to transfer data between host and graphics memory at up to 16 GB/s, while the graphics memory of the card can achieve a throughput of 336.5 GB/s.
However, on recent GPU architectures, interconnects such as Nvidia's NVLink~\cite{pascalarchitecturewhitepaper} and AMD's Infinity Fabric~\cite{vegawhitepaper} achieve throughputs of up to 300 GB/s, alleviating the impact of data transfers.

High performance on a GPU requires the given task to be expressed in a way that fits the GPU's hardware properties.
The main requirements are a large degree of parallelism and the possibility to achieve coalesced memory access as well as a common control flow among the threads within a warp.
Thus, memory-intensive tasks with complex data dependencies are typically difficult to execute efficiently on GPUs~\cite{yiao2010power,burtscher2012quantitative}.

Compared to many-core CPUs, programming for \acp{GPU} still requires profound knowledge of the GPU architecture~\cite{wkas2016gpgpu}.
As with many-core CPUs, the large number of configurable parameters render the performance tuning of GPU programs an important but challenging task~\cite{Torres-2011}.

\subsection{\acfp{APU}}
\begin{figure}[t]
	\begin{center}
		\includegraphics[width=4.5cm]{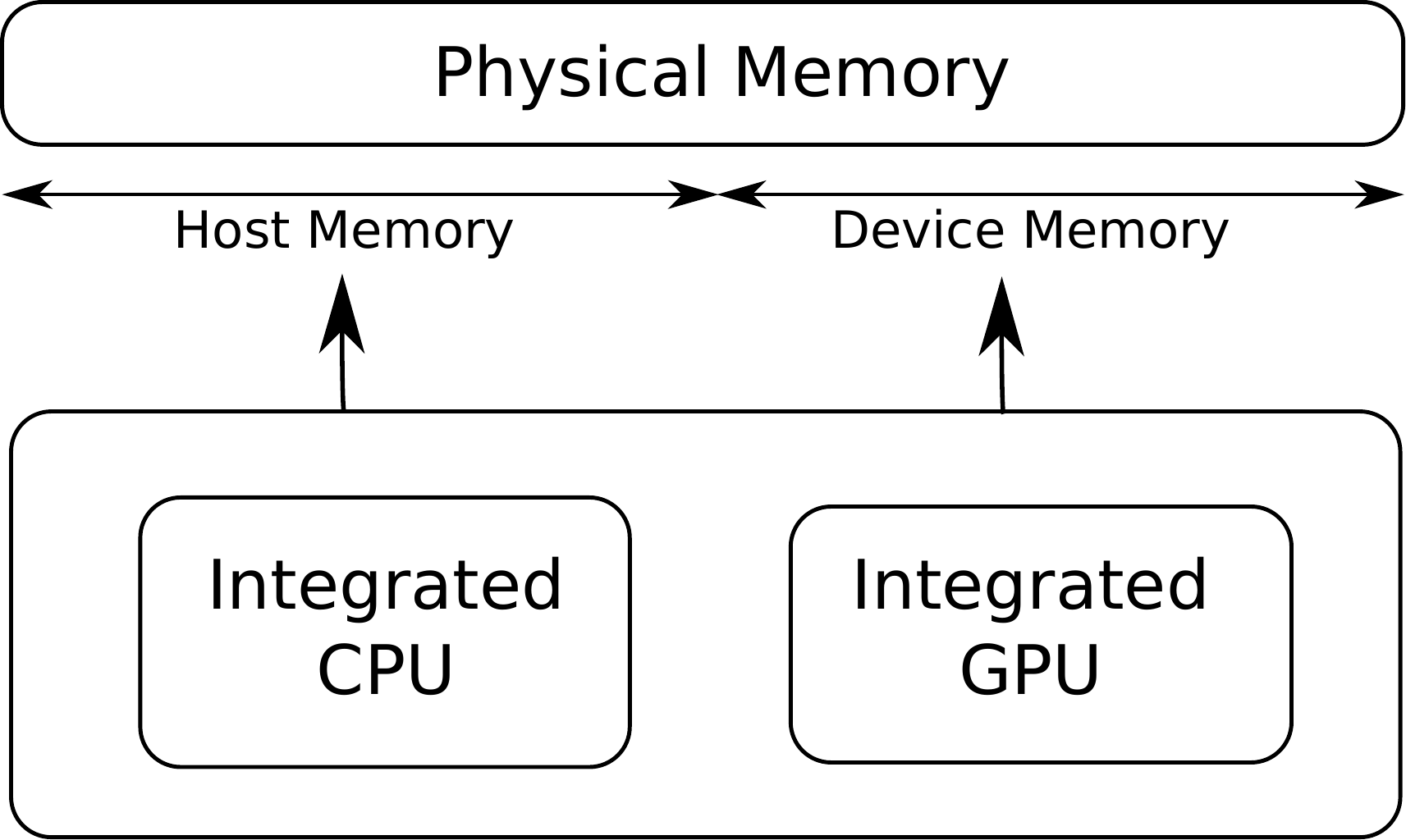}
		\caption{In an APU, memory is shared between the fused CPU and GPU.}
		\label{fig:apuarch}
	\end{center}
\end{figure}

\textbf{Architecture:}
\acp{APU} integrate CPU and GPU on a single die.
Although the term \acp{APU} has been coined by AMD, recent Intel CPUs with Intel HD Graphics follow a similar architecture.
Unlike stand-alone \acp{GPU}, the fused GPU of an \ac{APU} has direct access to the host memory through a low-latency and high-bandwidth bus.
Figure~\ref{fig:apuarch} sketches the high-level architecture of an \ac{APU}.

\textbf{Benefits:}
The main benefit of \acp{APU} is the opportunity for zero-copy memory access: since all memory is accessible both from the CPU and the GPU, costly data transfers over a relatively low-bandwidth bus like PCI-E can be avoided.
Zero-copy memory access also provides memory savings, as only one copy of an object in memory is required.
Further, scattered memory accesses which could only be handled inefficiently by the GPU can instead be performed by the CPU.

\textbf{Limitations:}
Existing \ac{APU} products have focused more on energy efficiency than high performance.
They typically contain fewer processing units than stand-alone CPUs and GPUs of the same hardware generation. For example, the Ryzen 5 2400G APU by AMD has 704 Vega-based stream processors, while the stand-alone graphics card AMD RX Vega 64 has 4096 stream processors.
As a consequence, compared to high-end stand-alone CPUs and GPUs, their computational power is relatively low.
Still, as will be discussed in Section~\ref{sec:addressing_the_challenges}, some works have considered \acp{APU} for accelerating agent-based simulations.

\subsection{\acfp{FPGA}}
\textbf{Architecture:} 
A \acl{FPGA} is an integrated circuit made of an array of interconnected \acp{CLB}.
Additionally, \acp{FPGA} are equipped with input and output pads and \ac{DSP} blocks.
FPGAs often provide various communication interfaces such as PCI-E, UART, and Ethernet.

A \ac{CLB} consists of several slices (sometimes also called logic cells), each slice containing a set of storage elements and \acp{LUT}.
A LUT has a number of inputs and outputs as well as flip flops that store a mapping between possible inputs and outputs. 
The mapping between inputs and outputs is defined by the users~\cite{hauck2010reconfigurable}.
The number of slices is one of the most important benchmarks to determine the computational power of an FPGA and can range from several thousand to several million.
For instance, the XCVU37P Virtex UltraScale FPGA from Xilinx has 2,851,800 slices~\cite{xilinxXCVU37P}.
In addition, the FPGA may have access to several GB of off-chip DRAM.

To describe the logic to be placed on an FPGA, typically a \ac{HDL} is used.
The two most widely used \acp{HDL} are VHDL~\cite{Heinkel-2000} and Verilog~\cite{Palnitkar-2003}.
In recent years, there have been intensive efforts to enable High-Level Synthesis, i.e., to generate FGPA layouts directly from high-level programming languages such as C, C++, or Java.
Recently, Intel released a dedicated SDK to use OpenCL to program \acp{FPGA}~\cite{intelopencl}.

FPGAs are sometimes used as a prototyping tool for the development of \acp{ASIC}, which requires an extensive and costly design process.
To the best of our knowledge, there is no work so far that employs custom ASICs for ABS.
In the field of \ac{DES}, a number of works have considered offloading of specific simulation tasks to ASICs~\cite{reynolds1993design,cleary1995architecture,fujimoto1988design,berlin1993design,lynch2011hardware}.
Notably, some of the envisioned components were fabricated physically~\cite{berlin1993design}.
Since the works on ASICs have only limited relevance to the field of ABS, we exclude them from our survey.

\textbf{Benefits:}
Due to the flexibility and high energy efficiency of FPGAs, they are frequently used for computationally intensive and highly parallelisable tasks.
For instance, FPGAs can be three orders of magnitude faster than GPUs when conducting specialised tasks such as encrypting a single 64-bit block by the Data Encryption Standard (DES)~\cite{che2008accelerating}.
In contrast to CPUs or GPUs, on which data paths are fixed, FPGAs provide flexible and customised data paths~\cite{Rahman-2017}.  
In the past years, \acp{FPGA} have received more attention in the field of simulation, particularly in \acl{EDA}, since hardware designs can be naturally expressed as FPGA layouts.

\textbf{Limitations:}
As with \acp{GPU}, \acp{FPGA} are connected to a host CPU without direct access to system memory.
The resulting need for data transfers can reduce the potential for performance gains.

FPGAs are regarded as lacking in programmability when compared to CPUs and GPUs~\cite{falsafi2017fpgas,che2008accelerating}. Although recent efforts towards high-level synthesis alleviate this limitation, manual tuning is still necessary to achieve the best performance~\cite{nane2016survey,fujimoto2017research}.

Finally, \acp{FPGA} are configured for a specific task.
Since reconfiguration can require multiple hours~\cite{Zohouri-2016}, FPGAs do not facilitate development processes that require fast iteration.
This may limit the applicability of FPGAs in early phases of simulation model development, where changes to the simulation model frequently occur and require immediate feedback for evaluation.

\section{Agent-based Simulation}
\label{sec:abs}

Agent-based modelling and simulation (ABMS) is a common approach~\cite{north2007managing} used for evaluating complex systems in domains such as traffic, crowds, economics, information propagation, biology, etc.
In the following, we characterise the modelling approach and discuss the properties of ABS with execution on heterogeneous hardware in mind.

\subsection{Modelling Approach}
\label{sec:modelling_approach}
In ABS, the simulated entities are agents that perform actions autonomously and interact with other agents based on certain rules.
ABS typically follows a \emph{Sense-Think-Act} cycle (e.g.,~\cite{riley2003next}): in the \emph{Sense} stage, an agent detects and analyses its neighbours as well as the environment in which it resides.
In the \emph{Think} stage, an agent makes judgement based on the information collected during the Sense stage.
The update of states takes place in the \emph{Act} stage.
The simulation time is typically advanced in fixed time steps at which all agents update their states.
However, if a model requires agents to update their states at variable points in simulation time, time advancement using a \emph{discrete-event simulation} (DES) approach may be more appropriate. In DES, state updates are performed through events scheduled for execution at discrete points in simulation time.
The simulation proceeds by iteratively executing the earliest remaining event, potentially scheduling new events in the process.

Independent of the time advancement mechanism, the defining characteristic of ABS distinguishing it from other simulation techniques is the \emph{autonomy} of agents, i.e., ``agents are endowed with behaviours that allow them to make independent decisions''~\cite{macal2005tutorial}.
Since the focus of this survey paper is on ABS, we exclude simulation domains such as physics and chemistry, which usually consider sets of entities that are passively affected by their environment.
However, we do discuss a number of methods proposed outside of the ABS domain with direct applications to ABS, e.g., GPU-based priority queues in the context of DES.

\subsection{Computational Aspects}
\label{sec:abs:subsec:computational_aspects}
When considering models with complex decision-making and behaviour at large scales, ABS can be computationally intensive.
In addition, due to the stochastic nature of ABS, the simulation of a given scenario is usually repeated multiple times in order to generate meaningful results, further increasing computational demands~\cite{kofler2014sampo}.
However, the \emph{Sense-Think-Act} cycle described above provides ample opportunities for parallel execution.
Since the \emph{Sense} and \emph{Think} stages are performed on a per-agent basis and do not modify the simulation state, these stages can be executed in parallel across all agents.
To achieve a consistent view of the simulation state for all agents, the state changes performed in the \emph{Act} stage must then be propagated to other processing elements.

When parallelising across multiple traditional CPUs or CPU cores, each processing element can execute the state updates for a subset of agents.
A well-known challenge in parallel and distributed ABS lies in partitioning the simulation workload among the processing elements.
Generally, there are two dimensions according to which a simulation can be partitioned~\cite{nagel2001parallel}: \emph{domain decomposition} partitions according to the simulation space (e.g., different roads in a traffic simulation), while \emph{functional decomposition} partitions according to different models (e.g., different layers of the network stack in a computer network simulation).
High-quality partitionings are characterised by low amounts of workload imbalance and communication among the processing elements.
When targeting heterogeneous hardware environments, the partitioning problem is complicated by the differences in the suitability of each hardware device for certain types of computations.
Thus, to achieve high performance, a key challenge is to find a suitable \textbf{hardware assignment} of the simulation tasks according to characteristics such as the instruction mix and the available degree of parallelism.

Since typically, some communication between the partitions cannot be avoided, techniques for the \textbf{minimisation of data transfers} are required to reduce the performance impact of the communication (e.g.,~\cite{jang2006agent}).

On CPUs, the ABS performance benefits from long instruction pipelines, large caches and effective branch prediction.
Beyond traditional parallel and distributed simulation, many-core CPUs enable high degrees of parallelism while supporting unmodified x86 code.
The key difference between a CPU execution in a multi-core and many-core setting is the interconnect through which the CPUs communicate.
Since the architecture of each core still closely follows a traditional CPU core, no major code adaptations are required to execute the agent update logic efficiently.

In contrast, since both GPUs and FPGAs achieve highest performance with computational problems of a highly regular structure, another challenge of executing ABS on hardware accelerators lies in dealing with the \textbf{scattered memory accesses} resulting from the largely unpredictable runtime behaviour of the simulation.
Further, irregular control flows and fine-grained computations make it challenging to fully utilise high-performance many-core devices. Thus, methods for the \textbf{maximisation of parallelism} are required.
As an example, consider a model where the simulation space is represented by a rectangular grid of cells, each cell being occupied by at most one agent.
Here, a simple hardware assignment is a one-to-one mapping of arithmetic units to cells.
On a GPU, due to its heritage in highly regular data-parallel tasks on pixel values, such a hardware assignment tends to enable high cache locality, minimisation of memory transactions, and high utilisation of the arithmetic units.
In fact, prior to the general-purpose programmability of GPUs, a number of works proposed translating grid-based simulations to operations on graphic textures (e.g.,~\cite{harris2002physically}).
The Brook language developed at Stanford~\cite{Buck-2004} automates the translation process to graphics operations.
Similarly, there is a correspondence between the structure of an FPGA and cellular grids~\cite{vourkas2012fpga}.
The basic function of a circuit in an FPGA can be seen as analogous to the function of a cell in a cellular automaton.

\begin{figure}[t]
	\centering
	\includegraphics[width=.60\textwidth]{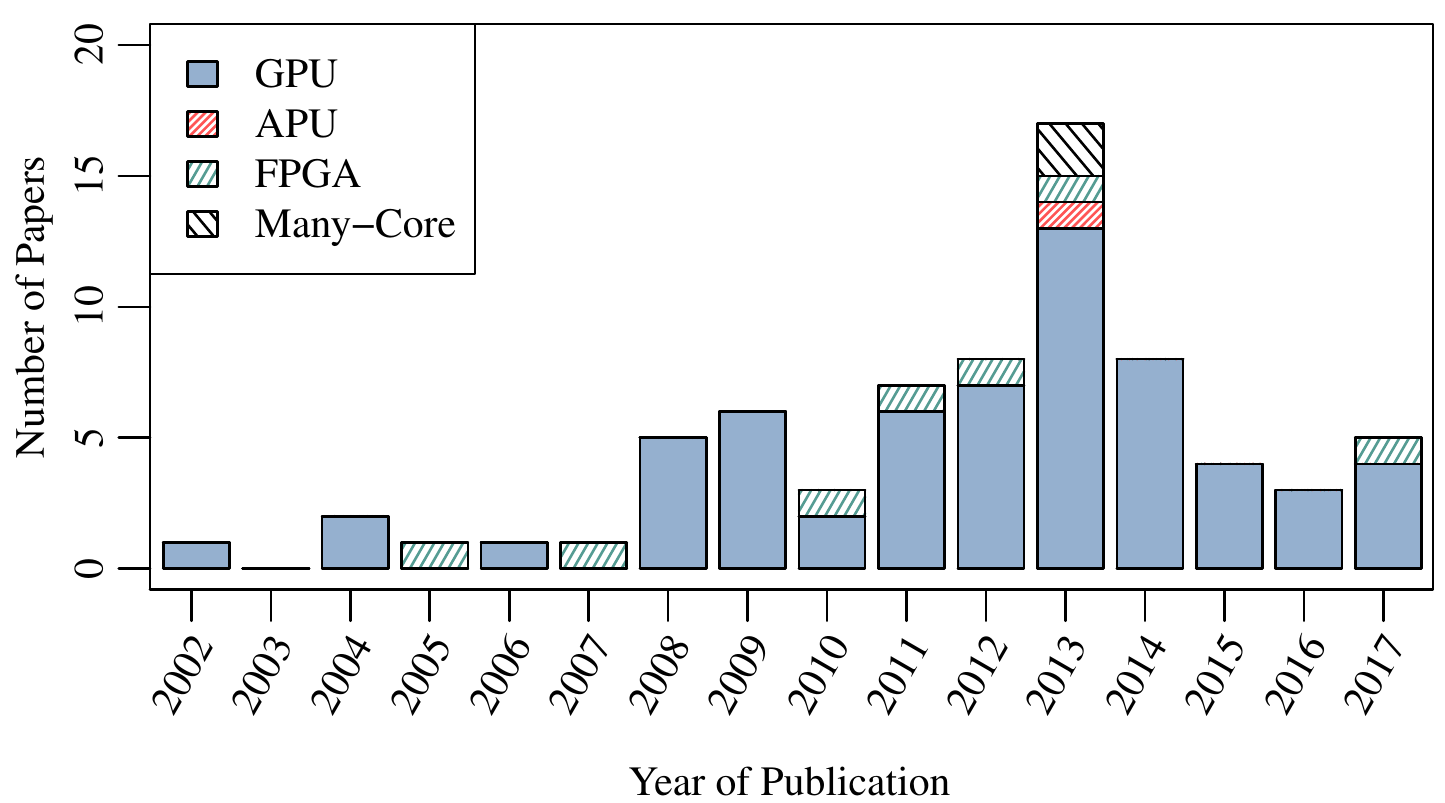}
	\caption{Publications on agent-based simulation on heterogeneous hardware by year and hardware type.}
	\label{fig:pubyear}
\end{figure}

However, in many models such as road traffic or social network simulations, the simulation space is a graph. Graph representations adapted to the architectural properties of the available hardware are required to efficiently support sensing an agent's neighbourhood and updating the simulation state while fully exploiting the available hardware. The general trend in the literature is moving towards supporting increasingly irregular types of simulations on accelerators.

The vast majority of literature on ABS using hardware accelerators has focused on GPUs (see Figure~\ref{fig:pubyear} for an overview of the number of publications since 2002)
We identify three reasons for the popularity of GPUs as accelerators for ABS: first, they are comparatively inexpensive. Second, in the recent years, the ease of programming of GPUs is slowly approaching that of CPUs.
Third, well-established programming frameworks such as OpenCL enable the formulation of models in a less hardware-specific manner.

In comparison, the use of FPGAs poses substantial challenges to modellers: only comparatively costly high-end FPGAs run at clock rates close to GPUs.
Thus, enormous degrees of parallelism are required to match a GPU's performance.
Further, while there exist some frameworks enabling high-level programmability, the achievable performance is limited compared to a more low-level specification of the desired logic in a hardware description language such as VHDL or Verilog.
As with GPUs, there is a need for libraries and frameworks that provide a higher degree of \textbf{abstraction from hardware specifics}.
Finally, the long runtimes of synthesis steps to generate an FPGA layout make model development and adaptation a cumbersome process.
Nevertheless, some works consider the use of FPGAs for ABS with promising results~\cite{georgoudas2010fpga,vourkas2012fpga}.

\section{Addressing the Challenges of Agent-Based Simulation on Accelerators}
\label{sec:addressing_the_challenges}
In the following, we discuss the techniques from the literature applicable to the key challenges in ABS on accelerators as identified in Section~\ref{sec:abs}: \textbf{hardware assignment}, \textbf{data transfer overheads}, \textbf{scattered memory accesses}, \textbf{maximisation of parallelism}, and \textbf{abstraction from hardware specifics}.
Table~\ref{tbl:challenges} summarises the systematisation of knowledge presented in this survey.
It contains our classification of challenges, techniques, and publications, as well as the considered types of accelerators.
For the publications that considered specific simulation models, Table~\ref{tbl:domains} shows the simulation domains and hardware platforms, providing researchers with pointers to relevant works in their respective domain.

\setlength\extrarowheight{2pt}
\begin{table*}[b]
	\scriptsize
	\setlength{\tabcolsep}{3pt}
	
	\centering
	\resizebox{\textwidth}{!}{%
	\begin{tabular}{c|c|c}
		\hline
		\textbf{Challenge} & \textbf{Technique} & \textbf{Publications}\\
		\hline
		\multirow{3}{*}{\shortstack[c]{Hardware\\assignment}} & \multirow{2}{*}{Static assignment by type of computation}  & 
		Many-Core~\cite{lai2013accelerating}, GPU ~\cite{hirabayashi2012toward,pavlov2013multi, Andelfinger-2011b, Bauer-2011,Bilel-2012,xu2014mesoscopic,song2017supporting, hermellin2016defining,michel2013translating}\\
		& & \cite{hermellin2015gpu,hermellin2016toward,zhang2017understanding}, APU~\cite{wang2013accelerating}, FPGA~\cite{tripp2005metropolitan,Cui-2011,vourkas2012fpga,georgoudas2010fpga}\\
		
		&\cellcolor{gray!25} Dynamic assignment based on runtime measurements & 
		\cellcolor{gray!25}GPU~\cite{belviranli2013dynamic,wen2014smart,grasso2013automatic,kofler2013automatic,zhang2017understanding,grosser2016polly}, FPGA~\cite{belviranli2013dynamic}\\
		\hline
		\rule{0pt}{2ex}		\multirow{3}{*}{\vspace{+0.3cm}\shortstack[c]{Data transfer\\overheads}} & Overlapping of communication and computation & 
		GPU~\cite{Kunz-2012,Bauer-2011,Bauer-2014}\\
		& \cellcolor{gray!25} Computation replication at partition boundaries &\cellcolor{gray!25} GPU~\cite{Aaby-2010,Zou-2013}\\
		\hline
		\multirow{4}{*}{\vspace{0.15cm}Scattered} & Manual caching in shared memory & 
		GPU~\cite{richmond2009cellular,Zou-2013,Li-2014}\\
		& \cellcolor{gray!25}Heuristics for agent update order & \cellcolor{gray!25} GPU~\cite{andelfinger2018fastforwarding,Harris-2012,Jin-2012,Jin-2013}\\
		memory accesses 	&& APU~\cite{wang2013accelerating}, GPU~\cite{harris2002physically,lysenko2008framework,Perumalla-2008a,Perumalla-2008b,Perumalla-2009,kolb2004hardware,Richmond-2008,strippgen2009using,vigueras2014accelerating,d2009data}\\
		& \multirow{-2}{*}{\shortstack[c]{Representation of irregular data structures\\by arrays and grids}} & 
		\cite{Park-2010,Zhen-2014,Seok-2012,kofler2014sampo,Swenson-2015,Andelfinger-2014b,Liu-2017,Park-2011,Wenjie-2013,baudis2017performance},
		FPGA~\cite{Rahman-2017,model2007discrete}\\
		\hline
		\multirow{3}{*}{\vspace{-0.4cm}\shortstack[c]{Maximisation of\\parallelism}} & Multiple replications in parallel & 
		GPU~\cite{pavlov2013multi,shen2011agent,laville2013using,Kunz-2012,Li-2015,yoginath2018}
		\\
		& \cellcolor{gray!25 }Window-based event execution &\cellcolor{gray!25} 
		GPU~\cite{Perumalla-2006,Buck-2004,Park-2008, Park-2011, Sang-2013,Andelfinger-2014b,Zhen-2014,Tang-2013b}\\
		& Speculative execution &GPU~\cite{Li-2013, Liu-2017}, FPGA~\cite{model2007discrete}\\
		& \cellcolor{gray!25} Computation sorting &\cellcolor{gray!25} GPU~\cite{Kunz-2012,Tang-2013b,kofler2014sampo}\\
		\hline
		\multirow{2}{*}{\shortstack[c]{Abstraction from\\hardware specifics}} & Frameworks to support simulation development & Many-Core~\cite{laville2013mcmas}, GPU~\cite{richmond2009cellular,Richmond-2011,laville2013mcmas,lysenko2008framework,Heywood-2015}\\
		& \cellcolor{gray!25} Unified memory access &\cellcolor{gray!25} GPU~\cite{lee2009openmp,yan2009jcuda,jablin2011automatic,jablin2012dynamically}\\
		\hline
	\end{tabular}
	}
	\caption{A classification of the challenges in agent-based simulation on accelerators along the relevant works addressing them.}
	\label{tbl:challenges}
\end{table*}
\setlength\extrarowheight{0pt}

\begin{table*}[t]
	\scriptsize
	\centering
	\begin{tabular*}{\textwidth}{c|c|c|c|c}
		\hline
		\textbf{Domain/Hardware}	& \textbf{Many-Core CPU} & \textbf{GPU} & \textbf{APU} & \textbf{FPGA}\\
		\hline
		Mobility  & & ~\cite{Perumalla-2008b} ~\cite{Perumalla-2009}~\cite{strippgen2009using}  ~\cite{shen2011agent} ~\cite{wang2012gpu}&&\\ & & ~\cite{xu2014mesoscopic}~\cite{song2017supporting}~\cite{Heywood-2015}~\cite{hirabayashi2012toward} ~\cite{andelfinger2018fastforwarding}& ~\cite{wang2013accelerating} & ~\cite{tripp2005metropolitan}\\
		\hline
		Biology  & ~\cite{laville2013mcmas} & ~\cite{Richmond-2008} ~\cite{d2009data} ~\cite{Aaby-2010} ~\cite{Perumalla-2008a}~\cite{richmond2009cellular}&&\\ & &~\cite{Zou-2013} ~\cite{Tang-2013b} ~\cite{hermellin2015gpu} ~\cite{hermellin2016toward} ~\cite{kofler2014sampo} ~\cite{Richmond-2011}~\cite{Li-2014} & & ~\cite{Cui-2011}\\
		\hline
		Ecology  &  & ~\cite{laville2013using}~\cite{yoginath2018} & & ~\cite{vourkas2012fpga} \\	
		\hline
		Social  & ~\cite{lai2013accelerating} & ~\cite{Jin-2012} ~\cite{Jin-2013} ~\cite{Zhen-2014} ~\cite{vigueras2014accelerating} ~\cite{Li-2015} ~\cite{wkas2016gpgpu} ~\cite{Li-2013} & &~\cite{georgoudas2010fpga} \\
		\hline
		Physics and Chemistry  & &  ~\cite{kolb2004hardware} ~\cite{Seok-2012} ~\cite{michel2013translating}~\cite{Bauer-2014} ~\cite{Perumalla-2006}~\cite{harris2002physically}~\cite{yoginath2018} & & ~\cite{model2007discrete} \\
		\hline
		Network  & & ~\cite{Kunz-2012} ~\cite{Bilel-2012} ~\cite{Andelfinger-2011b} ~\cite{Park-2008} ~\cite{Park-2011} ~\cite{Sang-2013} ~\cite{Tang-2013b} ~\cite{Andelfinger-2014b}& &  \\
		\hline
	\end{tabular*}
	\caption{Simulation model domains considered in the works covered in the survey.}
	\label{tbl:domains}
\end{table*}

\subsection{Hardware assignment}
\label{subsec:hardware_assignment}

One of the main challenges in parallel and distributed computations in heterogeneous hardware environments lies in finding a suitable partitioning, i.e., assignment of a given problem to the available hardware~\cite{fujimoto2017computational}.
We discuss techniques that have been used to address this problem according to two different, yet interrelated, aspects: first, we consider techniques to select suitable hardware for sub-tasks according to their ability to efficiently execute certain types of computations.
The minimisation of data transfers among the partitions running on separate devices will be considered in the next subsection.
The existing approaches can be roughly categorized as follows:

\begin{enumerate}[label=\textbf{\arabic*}.]
\item \textbf{Static assignment}: if the simulation model involves different types of computations that clearly suggest a certain hardware mapping, it may be sufficient to partition the model prior to a simulation run without any adaptation during runtime.
For instance, model segments involving large numbers of independent floating point operations may be well-suited for execution on a GPU, whereas segments with highly data-dependent control flow suggest the execution on a CPU.

\item \textbf{Dynamic assignment}: frequently, the dynamic behaviour of a simulated system at runtime translates to unpredictable computational patterns.
In such cases, maintaining high performance may require an adaptation of the hardware mapping based on performance measurements at runtime.
An inherent challenge of dynamic assignment is the trade-off between the performance increase through an improved assignment and the costs of runtime measurements and re\hyp{}assignment.
\end{enumerate}

An ample body of research has considered the parallelisation of general programs onto heterogeneous platforms, which is an enormous challenge due to the arbitrary control flows and memory access patterns that can be present in general programs.
Thus, typically, the approaches limit themselves to program portions that are particularly amenable to parallelisation on accelerators.
In the case of ABS, constraints such as the separation of data into a per-agent state and the limited sensing range of agents somewhat simplify the problem of parallelisation, potentially enabling a higher degree of automation in the hardware mapping.
In Section~\ref{sec:towards_an_automated_offloading_procedure}, we outline the vision of an automated approach and the required building blocks towards an automated hardware mapping for heterogeneous ABS.

\subsubsection{Static assignment}
Several authors have compared approaches to statically assign portions of the simulation workload to an accelerator.
Hirabayashi et al.~\cite{hirabayashi2012toward} compare a fully GPU-based execution to a hybrid CPU-GPU scheme where the CPU controls the progress of the simulation and calls the GPU for specific tasks.
In a traffic simulation based on the Optimal Velocity model~\cite{bando1995phenomenological}, the fully GPU-based acceleration clearly outperforms the hybrid scheme, although the lack of synchronisation operations across blocks introduces errors into the simulation results.

A similar categorisation is presented by Pavlov and M{\"u}ller~\cite{pavlov2013multi}, who conclude that a CPU-GPU approach where both the CPU and the GPU hold duplicated or partial agent and environment information is the most promising. 

\begin{figure}[b]
	\centering
	\begin{subfigure}{0.49\textwidth}
		\centering
		\includegraphics[width=1.0\linewidth]{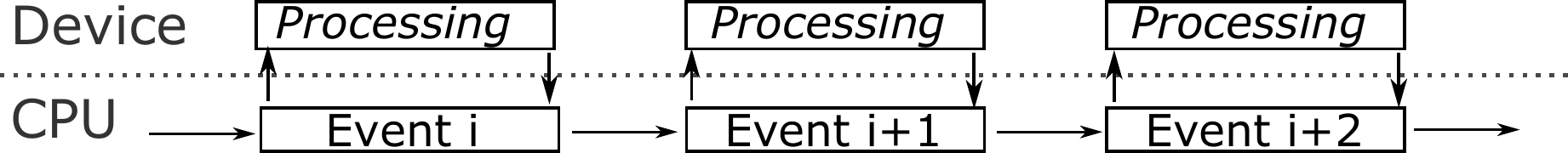}
		\caption{Hybrid CPU/Device.}
		\label{fig:arch-philipp-hybrid}
        \vspace{0.3cm}
	\end{subfigure}
	\begin{subfigure}{0.49\textwidth}
		\centering
		\includegraphics[width=1.0\linewidth]{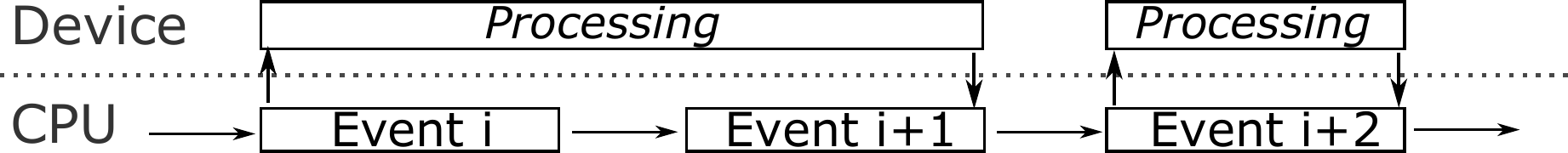}
		\caption{Event Aggregation.}
		\label{fig:arch-philipp-event_aggregation}
        \vspace{0.3cm}
	\end{subfigure}%

	\begin{subfigure}{0.49\textwidth}
		\centering
		\includegraphics[width=1.0\linewidth]{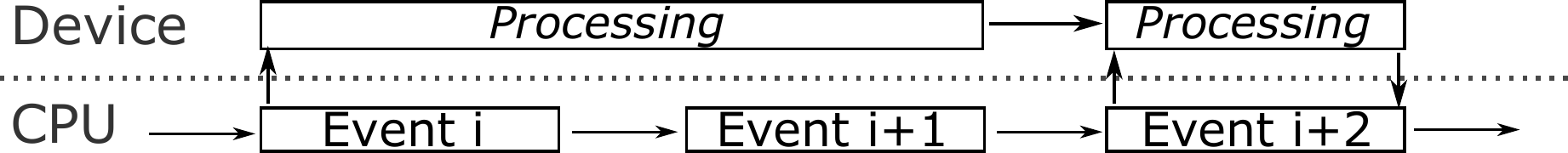}
		\caption{Memory Reuse.}
		\label{fig:arch-philipp-memory_reuse}
        \vspace{0.3cm}
	\end{subfigure}
	\begin{subfigure}{0.49\textwidth}
		\centering
		\includegraphics[width=1.0\linewidth]{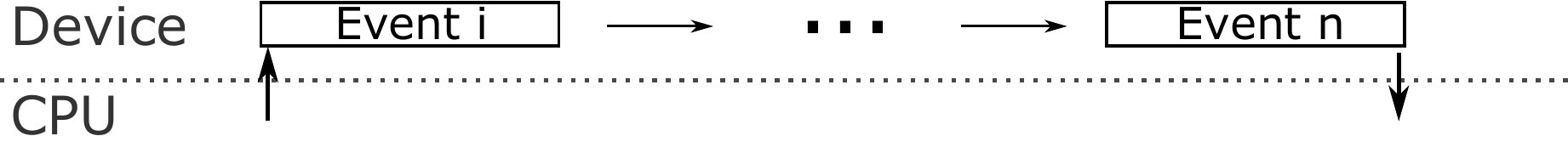}
		\caption{Fully Device-based Simulation.}
		\label{fig:arch-philipp-fully_gpu-based_simulation}
        \vspace{0.3cm}
	\end{subfigure}%
	\caption{Four CPU-Device simulation schemes~\cite{Andelfinger-2011b}. Devices in the figure can be GPUs or many-cores.}
	\label{fig:arch-philipp}
\end{figure}

Andelfinger et al.~\cite{Andelfinger-2011b} compare four GPU/CPU simulator architectures in the context of discrete-event simulations.
In a basic CPU/GPU hybrid scheme (cf.~Figure~\ref{fig:arch-philipp-hybrid}), the CPU offloads each event to the GPU individually.
Input data is transferred to the GPU at the beginning of the cycle.
After the computation is completed, the output data is transferred back to the CPU.
By aggregating independent events and executing them in parallel in a single step, data transfers are reduced (cf.~Figure~\ref{fig:arch-philipp-event_aggregation}).
A further reduction in data transfers is achieved by leaving computation results required by subsequent events in graphics memory (cf.~Figure~\ref{fig:arch-philipp-memory_reuse}).
Finally, if the entire simulation is ported to the GPU, data transfers are only required at the start of the simulation and once the simulation terminates (cf.~Figure~\ref{fig:arch-philipp-fully_gpu-based_simulation}).
While the simulation performance increases with each of the above optimisations, more and more changes to the simulator architecture are required, complicating development and reducing maintainability.

Two approaches for parallelisation are explored, corresponding to hybrid CPU/device (cf.~Figure~\ref{fig:arch-philipp-hybrid}) and fully device-based simulation (cf.~Figure~\ref{fig:arch-philipp-fully_gpu-based_simulation}) in~\cite{Andelfinger-2011b}, respectively.
Lai et al.~\cite{lai2013accelerating} implement the four geo-spatial applications of Kriging interpolation~\cite{shi2013kriging}, ISODATA~\cite{ball1965isodata}, Game of Life~\cite{gardner1970mathematical} and an urban sprawl simulation using cellular automata~\cite{wu2002calibration}. The authors compare the performance achieved when using one CPU per execution node, one GPU per node and 60 cores per CPU-based many-core accelerator, using MPI for inter-node communication in each instance.
The authors conclude that the use of GPUs and CPU-based many-core accelerators both provide a performance benefit over the purely CPU-based execution.
Given a sufficiently large number of assigned processors, using the CPU-based many-core accelerator with fully device-based simulation achieves similar performance as the GPU-based acceleration.


A number of authors considered hardware assignments tailored to specific simulation models.
For instance, when the underlying simulation can be clearly separated into model computation and management tasks, a master-worker scheduling approach as shown by Bilel et al.~\cite{Bilel-2012} in the context of large-scale mobile networks simulation can be used.
In the proposed design, the model is executed on the GPU, while the CPU orchestrates the event scheduling, simulation status monitoring, and memory allocation.
A node of the simulated network is partitioned into multiple processes, each process being executed by one GPU warp.

The nature of traffic simulation allows for a relatively straight-forward static hardware assignment according to different simulation aspects.
Xu et al.~\cite{xu2014mesoscopic} and Song et al.~\cite{song2017supporting} introduce a mesoscopic traffic simulation in a hybrid CPU-GPU architecture, assigning the agent mobility to the GPU, whereas the route calculation, the agent generation, and file reading and writing remain on the CPU.
The two parts run asynchronously to hide data transfer latencies.

Bauer et al.~\cite{Bauer-2008} consider a combined continuous-discrete simulation and assign the continuous part the GPU and the discrete part to the CPU.
The benchmark model PHOLD~\cite{fujimoto1990performance} is employed to explore different GPU configurations by varying the thread block size, the number of floating point instructions, the data transfer volume, and the communication pattern.
The authors conclude that while keeping the GPU fully utilised poses a challenge, models in a combined simulation with a large number of floating point computations can benefit from GPU acceleration.

Taking into account zero-copy memory access, Wang et al.~\cite{wang2013accelerating} show how a road traffic simulation can be accelerated using an APU.
In their simulator, sorting of agent states is required to locate each agent's neighbours.
While the APU's GPU resources perform state updates and local sorting, the sorting across GPU blocks is handled by the CPU resources.
The work separation can be carried out efficiently using zero-copy memory accesses.

In the ABS framework TurtleKit, the authors leave the simulation of agent behaviours to the CPU while environment dynamics are handled by the GPU~\cite{michel2013translating}.
With this approach, the authors aim to reduce the impact of the GPU acceleration facilities on the maintainability of the simulator code.
To increase the performance, portions of the agent behaviour that do not depend on the agent state, e.g., perception of properties of the environment, are performed on the GPU independently of individual agents for all locations and time steps.

Considering FPGAs, Tripp et al.~\cite{tripp2005metropolitan} showed how the movement of agents on individual lanes can be computed on the FPGA, while the agents' transitions from one road to another as well as the behaviour at intersections is computed on the CPU. 
However, most works on ABS on FPGAs focus on simulation models that allow for statically assigning the entire simulation to an FPGA.
For instance, the representation of cellular grids on FPGAs is explored by Vourkas and Sirakoulis~\cite{vourkas2012fpga}, who implement an environmental model simulation based on cellular automata (CA).
The authors note the structural similarity between a two-dimensional cellular automaton and an FPGA (cf.~Section~\ref{sec:abs:subsec:computational_aspects}).
A lattice of cells is simulated, each Configurable Logic Block (CLB) simulating one cell.
In case the number of cells exceeds the number of CLBs, the simulation lattice is partitioned into several layers, which are processed one after the other.
 Similarly, Cui et al.~achieve high performance with grid-based cellular automata on an FPGA~\cite{Cui-2011}.
A pipeline comprised of address generation, reading from memory, data alignment, rule computing, and updating of memory is applied to maximise throughput.
A similar method can be applied to cellular automaton-based crowd evacuation models as demonstrated by Georgoudas et al.~\cite{georgoudas2010fpga}. 

General guidelines for development of GPU-accelerated ABS starting from a CPU-based simulator implementation are proposed in~\cite{michel2013translating,hermellin2016defining}.
Their methodology requires the decomposition of the simulation model into small task modules and the heuristic identification of modules suitable for execution on a GPU.
As a heuristic, the authors state that loops and code segments with low amounts of conditional branching tend to be suited for execution on a GPU.
Then, the original task modules are manually replaced with GPU-executable modules.
Several case studies~\cite{michel2013translating,hermellin2015gpu,hermellin2016toward} show promising speedup by deploying this method.

Generally, the approaches relying on static hardware assignment split the simulation workload into coarse-grained functional tasks so that some tasks are clearly suited for a certain hardware device. To minimise trial-and-error, heuristics may be applied to identify a suitable mapping of tasks to the hardware. For instance, in the literature, tasks involving large numbers of parallel floating point operations are among the most common tasks offloaded to accelerators. Further observations are made by Zhang et al.~in the context of co-running programs on a CPU and GPU or in an APU~\cite{zhang2017understanding}: 1.~programs that are suitable to run in a CPU/GPU environment tend to have low memory bandwidth usage, 2.~most programs suitable for the APU allow for a large amount of overlap between CPU and GPU computations.

\subsubsection{Dynamic assignment}
\label{subsubsec:dynamic_assignment}
While a wide range of literature has considered the problem of dynamically adapting a partitioning of agent-based simulations to multiple CPUs (e.g.,~\cite{cosenza2011distributed,long2011agent,Xu-2017a}), we are not aware of such works that specifically target heterogeneous hardware environments.
In the following, we outline recent works on dynamic assignment of general computational workloads to heterogeneous hardware.
Since these works are generic, they cannot rely on knowledge of the general structure of ABS simulators or on model knowledge.

Belviranli et al.~\cite{belviranli2013dynamic} propose a self-scheduling scheme for partitioning generic application workloads into blocks and assigning them to CPUs, GPUs and FPGAs.
The proposed system consists of two phases: 
in the first phase, the system performs an online training with a small amount of data to estimate the maximum workload capacity size of each hardware device.
Fast convergence is achieved by fitting four sampled data points to a logarithmic function.
Once the capacity is determined, the processing unit's performance can be inferred from the same data.
When the change of processing speed between two samples drop below a threshold, it is used as the final estimated value.
In the second phase, the remaining workload is partitioned based on the percentage of the individual processing speed to the total speed of all available processing units, enabling faster processing units to handle a larger portion of the workload.

Some authors use machine learning techniques such as support vector machines, artificial neural networks and decision trees to distribute the workload of OpenCL programs to CPUs and GPUs.
For example, Grasso et al.~\cite{grasso2013automatic,kofler2013automatic} and Zhang et al.~\cite{zhang2017understanding} translate a single-device OpenCL program to a multiple-device program, while Wen et al.~\cite{wen2014smart} focus on scheduling multiple OpenCL functions to run in parallel on CPU/GPU. They train a machine learning algorithm according to a set of typical OpenCL programs and benchmarks.
The prediction generated by the machine learning algorithm guides the assignment of a portion of the computation to CPU or GPU. Their results show that the above three machine learning approaches outperform purely CPU- or GPU-based approaches. The scheduling scheme by Wen et al.~achieves a performance improvement compared to a first-come, first-served scheme and a scheme where computation-heavy task are handled by the GPU.

To automate the compilation of sequential programs for parallelised execution on heterogeneous hardware, Grosser and Groesslinger~\cite{grosser2016polly} present a compiler that generates CPU and GPU code.
Regions with mostly static control flow and sufficient computational intensity are detected and transformed to a formal representation to facilitate program transformations~\cite{grosser2012polly}. After optimisations have been performed to increase memory access locality and parallelism, CUDA code for GPU is generated from the formal representation.
A runtime library eliminates repeated memory allocations and unnecessary data transfers between CPU and GPU.
The decision whether a region is compute-intensive enough for execution on the GPU is made statically or at runtime using heuristics based on metrics such as the number of instructions. The authors conclude that the compiler is able to translate CPU code into cross-platform code with no performance penalty. For some computations, such as the correlation benchmark from polybench~\cite{polybenchc3.2}, significant speedup of up to two orders of magnitude can be achieved.   

The main difficulty in automated hardware mapping lies in determining the control flow and data dependencies of the original program. Current approaches either rely on the program code being formulated in languages such as OpenCL that express independent control flows explicitly, or only consider specific portions of programs such as loops with largely static control flow. In ABS, however, most of the available parallelism may exist across the update routines of separate agents. Thus, without semantic information describing the code structure, automatic detection of the parallelism is challenging. In Section~\ref{sec:towards_an_automated_offloading_procedure}, we sketch how the common structure of many ABS may be utilised to support the extraction of parallelism.

\subsection{Minimisation of data transfer overheads}
\label{subsec:minimisation_of_data_transfer_overheads}
Since most hardware accelerators are equipped with their own memory, simulations making use of accelerators typically require data transfers between host and accelerator memory.
Even with a high-quality partitioning of the simulation, these data transfers incur an overhead that reduces the speedup gained from the distributed computation.
In this section, we survey works that focus on minimising the cost of such data transfers.
The existing approaches can be roughly categorized according to the following techniques:

\begin{enumerate}[label=\textbf{\arabic*}.]
\item \textbf{Overlapping of communication and computation:} since some communication overhead between the processing elements involved in a simulation cannot be avoided, some authors proposed techniques to hide communication overheads by transferring data while independent computations are performed. Sometimes, the technique has been referred to as \textit{latency hiding} (e.g.,~\cite{bruning1994latency}).
\item \textbf{Computation replication at partition boundaries:} another technique to address communication overheads is to increase the amount of computation performed before synchronisation between processing elements is required.
This is achieved by duplicating some computations on multiple processing elements, thus delaying the need to resolve data dependencies across processing elements.
\end{enumerate}

\subsubsection{Overlapping of communication and computation}

One way of mitigating the overhead from data transfers between the host and an accelerator is to execute computations at the same time as data is being transferred.
In the approach described by Kunz et al.~\cite{Kunz-2012}, event computations are overlapped with data transfers across the CPU-GPU boundary, thus hiding data transfer latencies in a pipelined fashion.
Since events from multiple simulation instances are considered concurrently, there are substantial opportunities for overlapping these steps.

Bauer et al.~\cite{Bauer-2011,Bauer-2014} propose a generic API to optimise the data transfer between global memory and shared memory of CUDA GPUs using so-called warp specialisation. The warps within one cooperative thread arrays are split into two groups:
\emph{Dedicated memory warps} are in charge of data transfer between the on-chip and off-chip memory.
\emph{Compute warps} process the data.
The approach improves performance over thread-level separation between communication and computation since separate warps can follow divergent control flows without any performance penalty.
While their general idea can be applied to other types of independent processing elements, the warp-based implementation is specific to GPUs.

\subsubsection{Computation replication at partition boundaries}

In time-stepped ABS, at model time $t$ each agent updates its state based on the states of its neighbours at time $t-1$. If the simulation is distributed across multiple processing elements, synchronisation and data transfers are required to provide this information at each time step. The associated communication latencies may make up a substantial portion of the simulation runtime. Thus, some authors have proposed methods to reduce synchronisation by replicating some computations on multiple processing elements, similarly to performance optimisations in numerical computing~\cite{ding2001ghost}.

\begin{figure}[t]
	\centering
	\includegraphics[width=.32\textwidth]{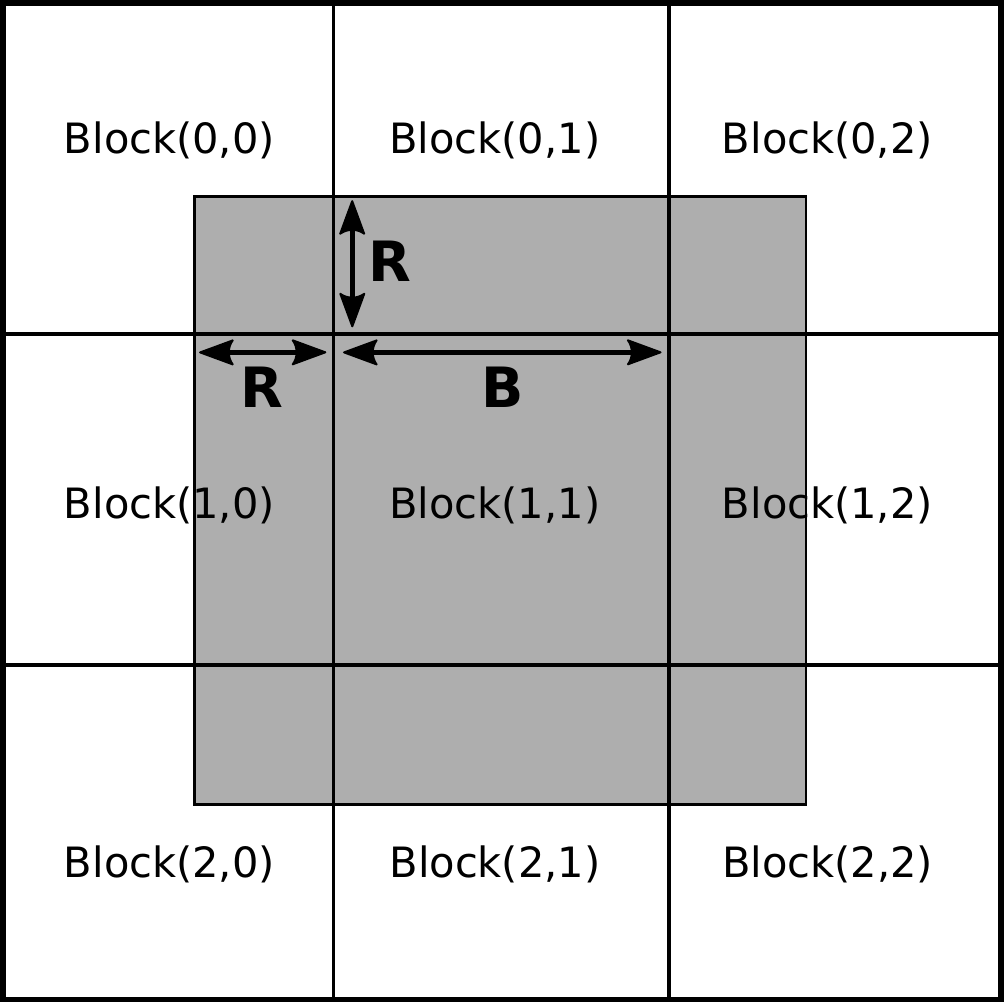}
	\caption{In the partitioning scheme by Aaby et al.~\cite{Aaby-2010}, $R$ cells are duplicated among neighbouring processing elements so that each processing element handles $(B+2R)\times(B+2R)$ cells.}
	\label{fig:aabyoverlap}
\end{figure}

Aaby et al.~\cite{Aaby-2010} present a multi-level data partitioning scheme for cellular simulations on multi-CPU/GPU clusters.
The simulation state is partitioned into blocks and each block is executed by a thread, a core, or a node, depending on the configured granularity.
In contrast to the traditional data partitioning into blocks of $B \times B$ cells and synchronisation at each time step, their approach partitions the data into several overlapping $(B+2R)\times(B+2R)$ blocks where $((B+2R)^2-B^2)$ cells form the overlapping area (cf.~Figure~\ref{fig:aabyoverlap}).
The computation in the overlapping area is performed redundantly by multiple processing units.
Thus, assuming that at each time step, a cell can only affect its immediate neighbours, $R$ time steps are required for a cell in the inner block to be affected by cells in another processing element.
Therefore, synchronisation is only required every $R$ time steps.
Between synchronisation points, an error propagates inwards within the overlapping areas, but does not affect the inner $B \times B$ cells before a new synchronisation occurs.
This partitioning approach is further employed in multi-GPU clusters on the node-, GPU-, block-, and thread-level, and for multi-CPU clusters at the node-, socket-, core-, and thread-level.

While Aaby et al.'s illustrates the idea based on cellular grids, the approach applies to general ABS.
The sensing range of agents is generally limited and provides an upper bound on the propagation of the effects of an agent's actions.
As long as overlapping segments of the simulation space can be distributed to the processing elements in a manner so that an effect requires at least $R > 1$ time steps, some synchronisations can be avoided. The generality of the approach is illustrated by Zou et al.~\cite{Zou-2013}, who extend the idea of computation replication to graph-based topologies in a GPU-accelerated epidemic ABS.

\subsection{Scattered memory accesses}
\label{subsec:scattered_memory_accesses}
Throughout the past decades, the increase in computational performance has outpaced the decrease in memory access latencies, leading to modern hardware designs towards ever-increasing cache sizes and deep memory hierarchies.

In the context of simulations, the issue of memory access latencies is particularly pressing: typically, a model's behaviour cannot be predicted before executing the simulation, significantly limiting the opportunities for a priori optimisation of data access patterns. However, commonalities between different simulation models can be exploited to propose data structures supporting efficient simulation of an entire range of models on a specific type of accelerator.

Since dynamic memory allocation on GPUs is costly~\cite{fujimoto2016research}, most GPU-based simulators allocate graphics memory for the main data structures such as the agent states statically (e.g.,~\cite{Li-2014}).
Another approach is to determine after each simulation step the required amount of memory and perform allocations accordingly~\cite{Park-2011}.

The existing approaches to address the issue of scattered memory accesses can be roughly categorized as follows:
\begin{enumerate}[label=\textbf{\arabic*}.]
\item \textbf{Manual caching in shared memory:} although the support for transparent caching has improved in recent years, achieving highest performance frequently still requires manual caching in low-latency segments of an accelerator's memory hierarchy.
Since typically the amount of low-latency memory is small, an iterative approach can be taken to limit the number of accesses to high-latency memory when accessing large amounts of data.
\item \textbf{Heuristics for agent update order:} since the data dependencies between agent state updates are typically not known prior to the execution of the simulation, minimising cache misses during the state updates is non-trivial.
Heuristic have been proposed, aiming to favour sequences of computations acting on the same agent data.
\item \textbf{Representation of irregular data structures by arrays and grids:} the hardware architecture of GPUs and FPGAs is designed so that highest performance is achieved when acting on regular data structures such as arrays and grids. Thus, efforts are taken to represent highly irregular data structures in a regular fashion. When covering the techniques from the literature, we first cover \emph{model-specific data structures} such as graph representations of a simulated road network. Subsequently, we discuss works covering two generic building blocks commonly required as part of ABS engines: \emph{priority queues and sorting}.
\end{enumerate}

\subsubsection{Manual caching in shared memory}

Richmond et al.~\cite{richmond2009cellular} propose to utilise the shared memory of the GPU as a manual cache.
In their agent-based simulation framework for cellular models in biology based on FLAME GPU~\cite{Coakley-2016}, they copy sets of messages to be transferred between agents into shared memory.
Each thread within a block can then efficiently iterate through the messages and identify those pertaining to the local agent.
Once all threads have iterated through the messages, the next sets of messages are loaded into shared memory.


Similarly, Zou et al.~\cite{Zou-2013} implement a manual software cache in shared memory to increase the performance of their graph-based epidemic simulation on GPU clusters. 
Before the simulation commences on the GPU, the CPU sorts the edges of the directed graph by the source vertex. 
Each thread block's shared memory stores edges originating from one specific node.
Since each block processes only edges originating from this node, a cache hit rate of at least 50\% is ensured.

In agent-based simulation, agents often influence and are influenced by their direct neighbours.
This fact can be exploited when arranging the simulation data in memory, reducing high-latency memory accesses when updating agents.
Li et al.~\cite{Li-2014} propose such a method for GPU-based ABS:
Assuming a constant number of agents, each agent is assigned to a GPU thread and its state data is permanently kept in global memory.
The simulation space is partitioned into a grid of rectangles.
Once a search for the neighbours within a circle around an agent is required, a search rectangle that encloses the searching circle is created, so only agents inside the search rectangle have to be considered.
Two approaches to utilise the GPU's shared memory are proposed: in the first approach, one block manages the searching process for a chunk C of close-by agents.
Per-block shared-memory loads the data of the agent and the agent's neighbours.
Each agent in C has a high probability of being in the other agents' neighbourhoods, so that these agents can frequently be accessed through the current block's low-latency shared memory.
However, since the limited shared memory capacity allows only for small numbers of agents to be stored, it is still likely that some neighbours are managed by another block and thus have to be accessed through global memory.
In the second approach, the shared memory loads the data of agents located in the union of all search rectangles of the agents' handled by the current block.
If the shared memory is not sufficient to hold all agents' data, the data is loaded as a sequence of chunks.
Of course, the increase in the search space given by the union of search rectangles leads to a higher number of unnecessary agent access through shared memory.
To address this problem, the union rectangle can be constructed on the warp level instead of the block level.

\subsubsection{Heuristics for agent update order}
The order in which agent updates are performed must adhere to the causal dependencies between the agent states and behaviours, e.g., in road traffic simulation, vehicles in direct proximity must be at the same point in simulated time to be able to interact according to the model specification.

Typically, this is achieved by a strictly time-stepped scheme in which agents always reside at the same time step, after which conflicts in the resulting agent states are resolved~\cite{yang2018evaluation}.
However, since in a typical simulation not all agents interact at each point in time, some agents may be updated further into the simulated future than others without affecting the simulation results~\cite{andelfinger2018fastforwarding}.
Harris and Scheutz have shown that distributed agent-based simulations can be accelerated by favouring agent updates that resolve dependencies across multiple processing elements~\cite{Harris-2012}.
This way, processing elements waiting for others to proceed can be unblocked, decreasing the amount of idle time.
Their approach can be applied independently of the underlying hardware platform, but requires bounds on the agent movement per time step.

Jin et al.~\cite{Jin-2012} present an information propagation simulation supporting execution on HPC systems and single GPUs and extend it to run on multiple GPUs~\cite{Jin-2013}.
Their focus lies on maximising the cache hit rate when traversing a graph according to rules defined by the simulation models.
Two categories of approaches are developed for the cascade model~\cite{goldenberg2001talk} and the threshold model~\cite{granovetter1978threshold}, which both simulate the propagation of information among nodes in a graph: vertex-oriented processing and edge-oriented processing.
For the vertex-oriented approach, the authors further describe two agent update orders: one iterates starting from active vertices, i.e., those that already have the information, and the other from inactive vertices.
Since the costs depend on the portion of active nodes, the simulation can switch dynamically between the two vertex-oriented approaches.
Finally, the edge-oriented approach iterates over the connecting edges between two vertices.
Since the number of edges is constant over a simulation run, the cost of the edge-oriented approach is less variable than that of the vertex-oriented approaches.
The authors achieved the highest performance when dynamically switching between the two vertex-oriented approaches.

\subsubsection{Representation of irregular data structures by arrays and grids}

GPUs and FPGAs are particularly suited for operations on regularly structured data.
However, many model types specify topologies that are more naturally expressed in terms of irregular structures such as graphs.
Further, execution of the simulator core itself may require operations on irregular data structures.

A basic optimisation commonly applied in works on GPU-based computing to improve memory access patterns is the transformation of the data layout in memory from arrays of structures (AoS) to structures of arrays (SoA) (e.g., \cite{richmond2009cellular,strippgen2009using}).
Commonly, sequential programs represent data in an AoS representation.
Since AoS bundles the properties associated with each object in object-oriented programming, or the states of agents in agent-based simulations, it is a natural way to represent data within these paradigms.
However, with an AoS data layout, parallel operations on the same property across many objects results in scattered memory accesses.  
An SoA data layout bundles the same property across all objects, which can increase cache hits rates and opportunities for memory access coalescing, thus improving performance substantially.

Beyond this simple optimisation, the data representation can be specialised for a given model to further improve performance.
In the following, we give an overview of methods applicable to ABS to achieve high performance by translating irregular data structures to a more regular form.

\textbf{Model-specific data structures}

Early works on executing ABS using GPUs frequently focused on cellular grids and translated the required computations into the graphics processing domain.
In a pioneering work done by Harris et al.~\cite{harris2002physically}, GPU shaders are used for implementing computations on the RGBA values in a texture that holds the agents' states.
The same idea is employed by Lysenko et al.~\cite{lysenko2008framework}, Perumalla and Aaby~\cite{Perumalla-2008a}, and Kolb et al.~\cite{kolb2004hardware}.

Perumalla et al.~\cite{Perumalla-2008a} evaluate the performance of running agent-based simulation entirely on a GPU.
They ported the cellular models Mood Diffusion~\cite{neumann2000mood,hess2006mood}, Game of Life~\cite{gardner1970mathematical} and Schelling Segregation~\cite{schelling2006micromotives}.
Through the Open Graphics Library (OpenGL), individual agent states are mapped to pixel colour values.
The authors report a speedup of 15 to 40 compared to CPU-based sequential execution.
Kolb et al.~\cite{kolb2004hardware} develop a particle simulation and a GPU-based collision detection mechanism built on the authors' previous work~\cite{kolb2001volumetric}.
Similarly, Richmond et al.~\cite{Richmond-2008} utilise the GPU's texture processing ability and map agent states onto texture data.
To accelerate the neighbourhood detection, the simulation space is partitioned dynamically according to the agents' current states.
The algorithm to generate partitions is borrowed from the particle pinning problem in rigid body particles physics~\cite{harada2007real,green2010particle}.
Identification of the start and end of the partition boundary is performed similarly to the method described in~\cite{oat2008efficient}. Textures are used to represent the agents' states and vertex texture fetching enables the search for the start and end of the partition boundary by comparing the partition value to the previous agent's state. 

To enable traffic simulations on GPUs, Perumalla~\cite{Perumalla-2008b} (and Perumalla et Aaby~\cite{Perumalla-2009}) proposes to model the road network as a grid made up of cells.
A road network in Cartesian coordinates is translated to a grid representation overlaying the network: a cell in the grid is marked as occupied when an edge of the original road network starts in the cell, passes the cell or ends in the cell.
In graphics memory, the cells' properties such as turning probabilities and length are stored in texture buffers.
Simulation is carried out by performing operations on the texture buffers.

A different method for traffic simulation on GPUs is presented by Strippgen and Nagel~\cite{strippgen2009using}, who propose a queue-based approach using CUDA.
Each road is represented as a single first-in, first-out (FIFO) queue stored in memory in the form of a ring buffer.
With the ring buffer, insertion of a vehicle entering a road and removal of a vehicle exiting a road is achieved with constant time complexity.
Coalesced memory access can be achieved by processing adjacent roads using adjacent threads.
Since the vehicles' mobility is modelled by a fixed per-link velocity, their approach can be considered mesoscopic.
Behaviours such as overtaking or lane-changing are not modelled and would require random insertions and removals from the ring buffers, which are associated with linear time complexity.

Other domains in which agent-based simulations have been successfully ported to GPUs using model-specific data structures include collision detection~\cite{vigueras2014accelerating} and a simulation study of tuberculosis~\cite{d2009data}.
In the former, a grid is split into tiles and data at the boundary of the tiles is replicated so that a consecutive space is occupied in the global memory of the GPU.
In the latter, the authors propose to use a sorted array according to the liveness status of agents, so that the state of a new agent can be stored in a memory location previously occupied by one of the dead agents.

\textbf{Sorting and priority queues}

Full or partial sorting are frequently required in agent-based simulations, e.g., for neighbourhood discovery or to implement priority queues (PQ) if time advancement is performed in a discrete-event manner.
These operations can involve large amounts of data-dependent and scattered memory accesses and are therefore challenging to implement efficiently on hardware accelerators.
Since this operation can occupy a substantial portion of the simulation runtime~\cite{Roenngren-1997}, a number of works have focused on memory layouts and algorithms for sorting and priority queues on accelerators.

As building blocks for time advancement in a discrete-event fashion, parallel reduction 
and bitonic sorting
are commonly used in GPU- and FPGA-based simulation~\cite{Park-2010,wang2013accelerating,Zhen-2014,Seok-2012,kofler2014sampo}.
We discuss these two operations jointly due to their structural similarities.
In both cases, an input array is split into chunks, each chunk being handled by one thread.
At each cycle, the sorted arrays/minimum values of two threads are then merged to form a new input array.
Thus, at each cycle, the number of chunks and active threads is cut into half.
The algorithm iterates until only one thread is active, leaving a sorted array or the global minimum value, respectively.

%
  
He et al.~\cite{He-2012} propose a parallel heap-based PQ on GPU based on a previous CPU-based design~\cite{deo1992parallel}.
The data structure resembles a binary min-heap, but stores $r$ items per heap node.
Items are inserted and extracted in a joint bulk operation that inserts up to $k \leq 2r$ and extracts up to $r$ elements.
At any time the root node is guaranteed to hold the highest-priority elements, while elements of lower priority are gradually inserted into deeper levels of the tree over the course of multiple insert-extract operations.
Parallelism can be exploited across the sorting operations on the items within a tree node, across the nodes on one level of the tree, and by processing all even-numbered and odd-numbered levels of the tree in parallel.
The costs of the queue operations can be hidden by performing them in parallel with the processing of extracted items.

Similarly, the FPGA-based DES simulators by Rahman et al.~\cite{Rahman-2017} relies on a pipelined heap~\cite{bhagwan2000fast} for storing events.
In contrast to the parallel heap by He et al., the pipelined heap is designed to achieve near-constant access times, but does not provide bulk operations.

A number of works avoid the need for a global PQ holding all future events.
Instead, the set of events is considered jointly in an unsorted fashion~\cite{Swenson-2015}, split by model segment~\cite{Seok-2012} or simulated entity~\cite{Zhen-2014,Andelfinger-2014b,Liu-2017}, split according to a fixed policy~\cite{Park-2011,Wenjie-2013}, or split randomly~\cite{model2007discrete}.
To determine the events that can be executed without violating the simulation correctness, a parallel reduction is performed to determine the minimum timestamp among the events.

Baudis et al.~\cite{baudis2017performance} evaluate the performance of PQs on a GPU implemented as a single parallel heap or as a set of ring buffers, implicit binary heaps, and splay trees~\cite{sleator1985self} in the context of DES and path finding on grids.
Their results indicate that for up to about 500 elements per PQ, ring buffers achieve the highest performance.
At larger element counts, implicit heaps outperform the other approaches in their study.
Their results suggest that higher performance is achieved by relying on multiple PQs, one for each agent or set of agents, compared to a single PQ holding all events.

\subsection{Maximisation of Parallelism}
\label{subsec:maximisation_of_parallelism}

The limited predictability of how the state of a simulated system evolves over time translates not only to scattered memory accesses, but also to an irregular control flow, which can negatively affect performance in two ways: first, variations in the computational intensity among the model segments may leave some processing elements idle. Second, the single-instruction multiple-thread execution model of GPUs requires divergent operations within a warp to be serialised.

The existing techniques to maximise the parallelism of ABS using accelerators can be roughly categorized as follows:

\begin{enumerate}[label=\textbf{\arabic*}.]
\item \textbf{Multiple replications in parallel:} full utilisation of a massively parallel accelerator requires large numbers of computations that are independent and can thus be executed in parallel.
If a simulation involves a sequence of mostly dependent computations, the overheads for communication may outweigh the gains from parallelisation. Thus, techniques have been proposed to perform computations from multiple simulation runs in parallel.
\item \textbf{Window-based event execution:} in simulations involving a discrete-event mechanism, only a proper subset of the simulated entities may require an update at a certain point in simulation time. Multiple authors have proposed gathering events across a window in simulated time, and executing these events in parallel. In effect, this approach forces a discrete-event approach into a time-stepped execution. A key difference among the techniques lies in whether the simulation correctness is strictly maintained.
\item \textbf{Speculative execution:} as in general optimistic parallel and distributed simulation~\cite{Fujimoto-2000}, computations may be performed speculatively to improve hardware utilisation. A rollback mechanism is required to revert to a correct simulation state after erroneous computations.
\item \textbf{Computation sorting:} when assigning neighbouring threads of a GPU to individual agents or events, divergence occurs if required computations are inhomogeneous. Some authors have proposed sorting of computations to minimise divergence.
\end{enumerate}

\subsubsection{Multiple Replications in Parallel}
\label{subsec:maximisation_of_parallelism:subsubsec:multiple_replications_in_parallel}

If an individual simulation run does not provide sufficient parallelism to fully utilise the available hardware, a Multiple Replications in Parallel (MRIP) approach~\cite{pavlov2013multi} can be applied, as shown by Shen et al.~\cite{shen2011agent}: in their approach, multiple replications of a traffic simulation~\cite{wang2012gpu} are executed in parallel on a GPU. Thus, both the parallelism among agents as well as the parallelism across replications can be exploited.
Laville et al.~\cite{laville2013using} implement a multi-agent simulation of microorganisms in soil for CPU/GPU in OpenCL. Each GPU thread manages one agent and each block is responsible for one simulation instance so that multiple simulation instances can run concurrently on one graphics card.
The idea is applied to discrete-event simulations by Kunz et al.~\cite{Kunz-2012}, focusing on executing parameter studies comprised of multiple replications on a GPU.

In addition to exploiting the parallelism across replications, Li et al.~\cite{Li-2015} aim to avoid unnecessary redundant computations common to multiple replications. They  propose a cloning mechanism for ABS on the GPU: in an ensemble simulation run comprised of multiple simulation instances, the computations that are common to multiple instances are only performed once.
When the behaviour of an agent diverges between two simulation instances, a clone of the agent is created.
Since the agent may affect other agents, cloning is performed according to the propagation of the effects of the original change in agent behaviour.
Across cloned simulation instances, neighbour detection can be aggregated to improve the utilisation of the GPU resources.
The benefit of cloning is limited when simulation runs diverge strongly, e.g., across multiple runs of a stochastic simulation using different seeds for random number generation.
Recently, the cloning approach has been applied to large-scale cellular simulations on GPU clusters~\cite{yoginath2018}.

\subsubsection{Window-based event execution}
On a GPU, all threads in a warp execute the same sequence of instructions on different elements of data. If no input data is available for some of the threads within a warp, the hardware utilisation is reduced.
In ABS, this issue is particularly obvious when time advancement is performed in a discrete-event fashion to accommodate varying state update intervals among the agents.
Then, events are scattered along the time axis, i.e., the probability that many events share the same timestamp may be low.
Thus, a simple parallelisation across the events at a certain point in model time may be insufficient.
An approach to address this problem is to execute DES models in a time-stepped fashion: all events within a certain time interval are executed in parallel.
The lower bound of this time interval is usually referred to as Lower Bound on Time Stamp (LBTS), which is similar to Global Virtual Time in optimistically synchronised parallel and distributed simulation~\cite{jefferson1985virtual}.
With a sufficiently large time step size, hardware utilisation is increased.
However, since dependencies between events are not considered, the simulation results may differ from a sequential execution.

A study comparing the performance of time advancement mechanisms for simulations on the CPU and the GPU is presented by Perumalla~\cite{Perumalla-2006}. They study diffusion simulations running in a time-stepped, discrete-event, and hybrid fashion.
The GPU variant is implemented in the GPU programming language Brook~\cite{Buck-2004}.
While the GPU outperforms the CPU in the time-stepped variant, it does not perform as well as the discrete-event implementation on the CPU. However, high speedup is achieved using the hybrid approach, where at each cycle, the minimum gap between two events is used as a time step.
The simulation time then advances according to this time step.
Fishwick.~\cite{Park-2008, Park-2011} present a method for queuing network simulation that executes a DES model in a time-stepped fashion.
The simulation time advances according to a fixed time step size, but skips periods where no events occur.
All events within the current time step are executed in parallel without considering their potential dependencies.
Although the simulation results may be affected by their approach, the authors show that for a queueing network simulation, error bounds can be given.
Other works assume a minimum time delta between an event and its creation (\emph{lookahead}) to guarantee the correctness of the simulation results~\cite{Sang-2013,Andelfinger-2014b,Zhen-2014}.
If lookahead is available, a window can be determined within which events are independent, allowing for parallel execution without affecting the simulation correctness.

The current time window is extended dynamically in work by Tang and Yao~\cite{Tang-2013b} to allow more events to be executed in parallel.
After executing all events within the current window, their algorithm evaluates the first event in the event queue with a timestamp larger than the LBTS that can still safely be executed according to the lookahead.

\subsubsection{Speculative execution}

To maintain the correctness of the simulation results when executing in parallel on an accelerator, the simulator must consider the dependencies between state updates. In some of the approaches described above, a time window is determined where state updates cannot affect each other. If it is difficult to determine a time window of sufficient size to extract substantial parallelism, a speculative (also referred to as \emph{optimistic}) approach can be employed: state updates are performed without regard for correctness, and rolled back if errors are detected.

The possibility of speculative execution of simulations on FPGAs has been first demonstrated by Model and Herbordt~\cite{model2007discrete}. They make use of an event predictor, which predicts the interaction between two particles and generates new events accordingly. Events may later be cancelled as a consequence of a false prediction.

Targeting GPUs, Li et al.~\cite{Li-2013} present an execution model that avoids divergent control flow by speculative event execution.
In an initial step, all events that may occur in the simulation are created.
Subsequently, all events are executed in parallel.
A scanning process detects and revokes causally invalid event executions: if an event leaves the simulation in an incorrect state according to a model-specific criterion, the erroneous event and all events created by it are revoked recursively.

A more general approach for GPU-based discrete-event simulation is presented by Liu and Andelfinger~\cite{Liu-2017}.
An optimistic execution scheme based on the Time Warp algorithm~\cite{jefferson1985virtual} implemented in CUDA is shown to be beneficial at low event density in simulated time.
To support rollbacks in case of erroneous computations, the authors show how the default random number generator in CUDA can be reversed computationally without storing additional data.

\subsubsection{Computation sorting}

On a GPU, threads within a warp following divergent branches of the control flow are serialised. For instance, if some threads in a warp execute the body of an \texttt{if} statement, whereas others execute the body of the corresponding \texttt{else} statement, the two sets of threads perform their actions one after another. Some approaches attempt to arrange the assignment of computations to the available threads so that branch divergence is minimised.

In their DES engine on the GPU, Tang and Yao~\cite{Tang-2013b} sort events by type before execution, i.e., by the code associated with the event.

The idea is applied to GPU-based execution of multiple simulation instances at the same time by Kunz et al.~\cite{Kunz-2012} (cf.~Section~\ref{subsec:maximisation_of_parallelism:subsubsec:multiple_replications_in_parallel}). 
If the simulation instances do not diverge too strongly, many events of the same type are available across multiple instances, enabling efficient parallel execution.

Kofler et al.~apply computation sorting to their ABS of mosquitoes~\cite{kofler2014sampo}.
In their simulator, a one-to-one mapping between agents and threads is used.
Depending on their current state, agents may perform different operations, which can result in taking different control flow branches during the state updates.
Thus, to reduce divergence among threads within a warp, agents are sorted by their current state, so that the state updates of adjacent agents share the same control flow.

\subsection{Abstraction from hardware specifics}
\label{subsec:abstraction_from_hardware_specifics}
Compared with model development in CPU-based environments, development for accelerators can be cumbersome and error-prone.
To avoid the need for modellers to gain deep expertise in programming for specific accelerators, several frameworks have been proposed that enable the specification of parts of the model structure and behaviour in a hardware-agnostic fashion.
The approaches to avoid the need for modellers to consider low-level aspects of accelerators can be classified as follows:

\begin{enumerate}[label=\textbf{\arabic*}.]
\item \textbf{Frameworks to support simulation development:} some authors have proposed generating partial model code to be executed on accelerators from domain-specific languages or the reliance on a library of pre-defined implementations of common simulation tasks and models.
However, in these approaches, developing a full ABS will typically still require manual implementation work using a comparatively low-level languages such as CUDA. Further, workload partitioning and assignment to different hardware devices is currently not considered by these approaches.
\item \textbf{Unified memory access:} since in most cases, the CPU and hardware accelerators involved in a simulation operate on separate memory, resolving data dependencies may involve cumbersome explicit data transfers.
A number of authors have proposed techniques to transparently access data in programs executed on heterogeneous hardware.
\end{enumerate}

\subsubsection{Frameworks to support simulation development}
In the Flexible Large Scale Agent Modelling Environment (FLAME GPU)~\cite{richmond2009cellular,Richmond-2011}, agent states are specified using the state machine model X-Machine~\cite{eilenberg1974automata,holcombe1988x}.
Modellers define agent states in an XML-based format, while state transitions, i.e., the code segments describing the state updates, have to be manually specified as CUDA code. Generic facilities for exchanging messages between agents are provided by the framework.
Use cases of the FLAME GPU framework can be, e.g., traffic simulation~\cite{Heywood-2015}.

Another framework called Many-Core Multi-Agent System (MCMAS) for GPU and other many-core architectures is introduced in~\cite{laville2013mcmas}.
The framework provides a high-level Java interface to OpenCL code as well as a set of pre-defined data structures and functions called plugins.
To implement agent models, users either rely on plugins or define their own plugins as OpenCL code that can be called from Java code.
The authors state that unlike FLAME GPU, in which models are targeted exclusively to the framework, the models defined in MCMAS can be reused by other agent-based simulators.


While FLAME and MCMAS both reduce the implementation work required to develop agent-based simulations targeting accelerators, these frameworks do not provide guidance or automation in distributing the simulation workload to the available hardware. Thus, manual experimentation is required to determine a suitable hardware mapping.

\subsubsection{Unified memory access}
GPGPU frameworks such as OpenCL or CUDA require the user to either explicitly trigger data transfers between host and device memory, to explicitly select certain variables or memory regions for access from both CPU and GPU code~\cite{CUDA-2018}, or to annotate the program to manage data transfers~\cite{yan2009jcuda,lee2009openmp}. These manual steps complicate the development of agent-based simulations in heterogeneous environments.
Some works aim to improve on this situation by transparently transferring required data between host and graphics memory.
However, in languages based on C or C++, static alias analysis, i.e., determining which pointers refer to the same memory regions, is known to be undecidable~\cite{jablin2011automatic}.

Jablin et al.~\cite{jablin2011automatic,jablin2012dynamically} presented the first fully automated data management system based on compilation steps and a runtime library.
The developer formulates his program and GPU code as if all data resides in host memory and can be accessed both from the CPU and GPU.
The proposed approach instruments the code to track accesses to different memory regions using code instrumentation and trapping of system calls.
To avoid the need for static pointer analysis, memory accesses through pointers are tracked by the runtime library.
In addition to transparently handling data transfers, CPU-GPU communication is optimised during compile time by re-ordering the program flow to reduce the alternation between computations and data transfers.
Unnecessary data transfers are avoided by leaving data in the GPU memory until it is accessed from the host.

While the work of Jablin et al.\ could be applied to automate data transfers in heterogeneous ABS, the detection of parallelism is not covered. In Section~\ref{sec:towards_an_automated_offloading_procedure}, we sketch research directions towards automation in porting ABS to accelerators.

\section{Towards an automated offloading procedure}
\label{sec:towards_an_automated_offloading_procedure}

From the observations in the previous section, we can state that there is a vast range of techniques covering the main challenges of high-performance ABS on hardware accelerators.
However, there exist only few ABS frameworks that support such accelerators.
Since existing agent-based simulation and model implementations typically target purely CPU-based environments, there is a clear need for processes and tools to support the transition to an execution on accelerators.
More specifically, modellers and simulationists should be supported in the parallelisation and hardware mapping as much as possible.
While methodologies have been proposed to systematise the steps of porting a simulation to a GPU~\cite{michel2013translating,hermellin2016toward}, there is still a lack of automated tools to support this process.

The problem of automatic parallelisation of general programs is a broad and active field of research~\cite{fujimoto2016research}.
Substantial successes have been achieved with respect to parallelisation of computationally intensive loops with predictable and mostly static control flow~\cite{grosser2016polly}, whereas the extraction of parallelism across complex and irregular programs is still a largely manual process.
Common approaches include specifying software systems using formalisms that express parallelism explicitly~\cite{hoare1978communicating,lamport2002specifying,majeti2016automatic} or annotating programs with parallelisation hints~\cite{dagum1998openmp}.
In essence, these approaches provide the compiler or parallelisation middleware with a dependency graph of the statements or code blocks within the original program.

Fortunately, many agent-based simulators and models roughly follow a common set of properties that simplify the extraction of parallelism.
We identify the following constraints that can be leveraged to support the parallelisation process:
\begin{enumerate}
\item \textit{Time-stepped execution}: usually, the model time is advanced in fixed increments. At each time step, all agents update their states.
\item \textit{Two states per agent}: to decouple the simulation results from ordering agent updates, simulators commonly support storing each agent's old state at $t-1$ and the new state at $t$ separately. During an update from $t-1$ to $t$, only read accesses are performed to the agents' states and the environment state at $t-1$, and only write accesses to the states at $t$. Thus, within an update, there are no read-after-write dependencies across agents.
\item \textit{Sense-Think-Act cycle}: we assume that agent updates follow the well-known Sense-Think-Act cycle (cf.~Sec.~\ref{sec:modelling_approach}), with one such cycle per model. 
\end{enumerate}

\begin{figure*}[t]
	\centering
	\includegraphics[width=0.99\textwidth]{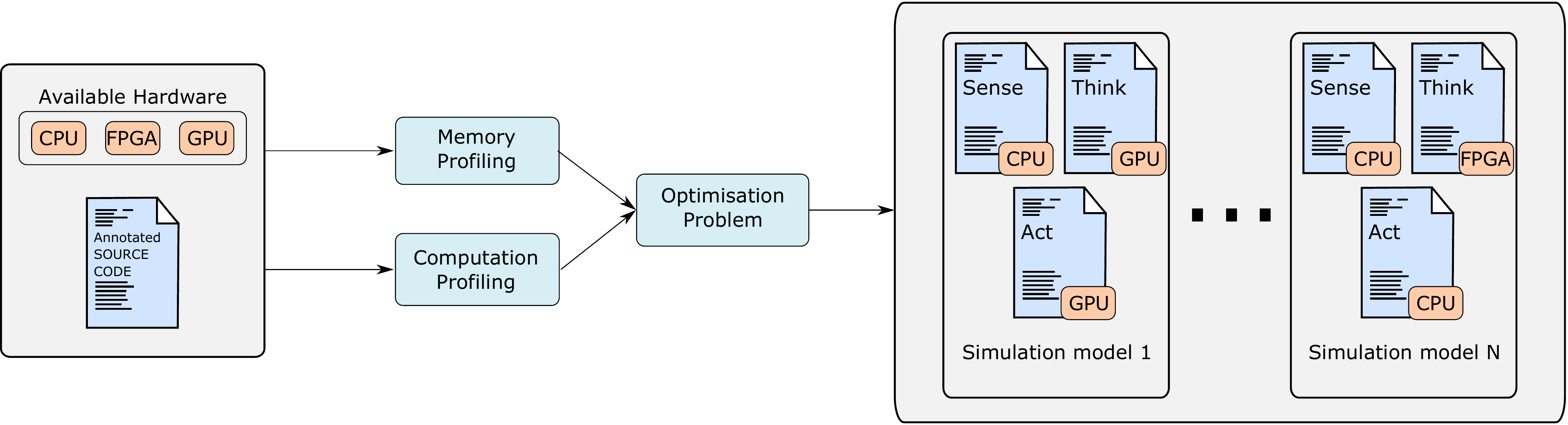}
	\caption{Workflow of the envisioned automated offloading procedure.}
	\label{fig:process}
\end{figure*}

With these constraints, a natural approach to parallelisation is to offload individual stages of a model's Sense-Think-Act cycle to an accelerator.
For instance, in crowd simulations using the social force model, the Think stage comprised of the computation of the force affecting an agent may be performed by one thread of a GPU per agent.

In the following, we sketch an envisioned workflow and the required tools to support users in porting an existing CPU-based ABS to a system equipped with hardware accelerators.
For the targeted simulator architecture, we assume a traditional \textit{master-worker} scheme, with the host CPU acting as the master and assigning work to the available accelerators at each time step.

\subsection{Proposed Work Flow}

The proposed semi-automated process is visualised in Figure~\ref{fig:process}.
To facilitate the automatic partitioning of the simulation source code into segments that can be outsourced to various types of hardware, we suggest manually annotating the source code according to the Sense-Think-Act paradigm.
From that it follows that the smallest unit that can be offloaded to a hardware accelerator in our proposed framework is one of these three stages.
Each of the stages is profiled in terms of memory and computational requirements.
According to the gathered requirements, an optimisation problem is solved to generate a hardware assignment (rightmost part of Figure~\ref{fig:process}).

For simplicity, we assume that all data required by the stage fits into one of the accelerator's memory entirely.
Otherwise, agents could be distributed across multiple accelerators or processed in batches, both implying additional communication costs.

\subsubsection{Input}

The source code is annotated manually to signify the stages of the Sense-Think-Act cycle, e.g., in the form \texttt{\#pragma sense\_begin}, \texttt{\#pragma sense\_end}, and so forth.
A simple example for a crowd simulation is given in Algorithm~\ref{alg:anno}.
In addition to the manual annotations, this clear separation may require refactoring of the simulation code.
By parsing the annotated source code, the framework obtains a mapping between code and stages that will later be enriched with data from measurements.

The second input is a specification of the available hardware.
Each hardware device is characterised by its available memory, computational performance, and host-device data transfer overhead. The computational performance can be stated in terms of single-threaded performance on CPUs, many-core CPUs, GPUs, and APUs.
We assume that for an FPGA, only model stages for which implementations already exist are eligible for offloading.
Thus, the computational performance of an FPGA is given with respect to specific model stages.  

\begin{algorithm}[t]
\footnotesize
\caption{Example for model code annotated with the stages of an agent update.}\label{alg:anno}
\begin{algorithmic}[1]
\State \texttt{\#pragma agent\_begin}
\vspace{0.1cm}
\State \textbf{class Agent:}
\Indent
\State Coord position;
\State \textbf{void  executeOnTimeStep():}
\Indent
\State  \texttt{\#pragma  sense\_begin}
\State  List agents = getNeighbouringAgents(position);
\State  \texttt{\#pragma sense\_end}
\vspace{0.1cm}
\State \texttt{\#pragma think\_begin}
\State  Coord velocity = computeVelocity(agents);
\State  \texttt{\#pragma  think\_end}
\vspace{0.1cm}
\State  \texttt{\#pragma  act\_begin}
\State  position = position + velocity;
\State \texttt{\#pragma  act\_end}
\EndIndent
\EndIndent
\vspace{0.1cm}
\State \texttt{\#pragma agent\_end}
\end{algorithmic}
\end{algorithm}

\subsubsection{Memory access profiling}

Now that the source code is partitioned into offloadable stages and the capabilities of all the hardware components are known, the data dependencies of each stage are determined.
Assuming a node in a graph represents one stage, then an edge in this graph represents a data dependency between these stages.
The dependency can refer to both agent or environment data.
The weight of the edge is the volume of the data that is accessed in the CPU-based simulator, i.e., that has to be transferred during offloading.
Usually, the Think stage only has a dependency on the Sense stage within the same model and agent (intra-agent dependency), whereas the Sense stage might depend on the environment and on other agents' states (inter-agent dependency).
Although we assume that an individual stage is not partitioned across multiple hardware devices, the amount of data gathered during the sense stage may vary over the course of the simulation.
For instance, if agents form clusters in the simulation space, the number of neighbours per agent may increase over time.
Thus, the data dependencies should be measured with respect to typical scenario conditions.
To avoid exceeding the memory capacity of one of the considered hardware devices, the profiling can be repeated for a worst-case scenario.

Tools exist that are able to ascribe memory accesses performed during a program run to the source functions, data structures or threads~\cite{ashraf2015memory}.
For instance, the tool PinComm constructs a dynamic data flow graph from instrumented program executions~\cite{heirman2010pincomm}.
The annotations shown in Algorithm~\ref{alg:anno} allow us to map function names to the separate agent update stages.
Thus, it is possible to obtain the amount of memory accessed within each stage.
Once the graph describing the amount of memory accesses across stages is created, the implications in terms of memory copying of moving a certain stage to a hardware device can directly be evaluated.
For example, if the Think stage is moved to the GPU and the Sense and Act stage remains on the host CPU, then the edges entering and leaving the Think node determine the data transfer overhead.
The actual cost of this copy procedure can be obtained from the device specification or through measurements.

\subsubsection{Computational profiling}

In addition to the memory requirements of each stage, information about the computational characteristics of each stage is required.
The estimated runtime could be inferred from hardware performance models~\cite{zhang2011quantitative,sim2012performance,chen2009first,baghsorkhi2010adaptive}.
Approaches as those described in Section~\ref{subsubsec:dynamic_assignment} can be applied to estimate the suitability of different agent update stages for execution on a certain accelerator.
By characterising the workload incurred by each stage in terms of instruction mix and memory accesses as well as the number of agents, the performance of executing the full-scale simulation can be estimated~\cite{grasso2013automatic,kofler2013automatic,zhang2017understanding,wen2014smart}.
Alternatively, if the runtime of a stage is dominated by a sub-task that can easily be ported to an accelerator, measurements with respect to this task can be performed directly on the accelerator~\cite{belviranli2013dynamic}.

\subsubsection{Optimisation problem}

Building on the graph that represents data dependencies, an optimisation problem of assigning stages to hardware types can be formulated, similar to the approach targeting embedded systems by Zhang et al.~\cite{zhang2017using}.
In essence, constraints are formulated so that each stage is assigned to the host or a device, resulting in an overall simulation schedule.
Importantly, the optimisation problem must reflect the data location after each stage or time step (e.g.,~\cite{majeti2016automatic}).
For instance, to avoid data transfers, it may be more efficient to execute two subsequent stages on the same accelerator.
The objective function of the optimisation problem is the overall runtime, i.e., the sum of all estimated execution times on the respective device and the incurred communication costs by distributing nodes of the dependency graph that are connected by an edge.

\subsubsection{Output}

The output of the optimisation steps is a recommendation of which stages should be executed on which hardware device.
It is then the task of the user to port the code of each stage so it can be executed on the assigned device.
This might require specific knowledge, e.g., programming in VHDL or OpenCL and can therefore be an obstacle to some researchers.
Given that some established simulation models are used by many researchers (e.g., a CSMA/CA model in network simulation or different car-following models in traffic simulation), a public repository of common simulation models could be created, similarly to the plugin approach used in MCMAS~\cite{laville2013mcmas}.
Researchers could download these crowd-sourced simulation models to enable parts of their own simulations to be run on heterogeneous hardware environments, and contribute their own model implementations.
Such a repository would also reduce the need to estimate execution times and improve the optimisation results by allowing direct measurements on the potential target devices.
Similarly, after porting a specific model stage, new measurements may be performed to provide the optimisation process with more accurate performance data.

\subsubsection{Discussion}

In our approach, we take a pragmatic perspective: while the envisioned workflow is achievable based on existing building blocks, our assumptions may leave substantial performance potentials unexplored. In particular, by assuming that models and their stages are both executed as a series of dependent steps, we only exploit the inter-agent parallelism within each stage, while any parallelism across stages is not considered. In the following, we revisit the key challenges of ABS using hardware accelerators and sketch techniques from the literature that could be applied to maximise the performance benefits given our assumptions.

The \textbf{hardware assignment} (cf.~Section~\ref{subsec:hardware_assignment}) is the main focus of the proposed work flow. Above, we describe a static assignment using a functional decomposition. Still, the optimisation problem that determines the hardware mapping could be updated according to runtime measurements.

To minimise \textbf{data transfer overheads} that cannot be avoided (cf.~Section~\ref{subsec:minimisation_of_data_transfer_overheads}), a bulk execution of multiple simulation runs would be feasible. The optimisation problem could be adapted so that the computational and memory requirements reflect those of each stage executed within multiple simulations runs at the same time. The output of the optimisation process would then be a schedule for an execution in a multiple replications in parallel (MRIP) fashion~\cite{pavlov2013multi,Kunz-2012} .

The technique of overlapping computations with data transfers seems challenging in our approach, since we assume a serialisation of the agent update stages. However, pre-fetching across stages may be performed by commencing data transfers once some agents have finished a stage.

\textbf{Scattered memory accesses} and the \textbf{maximisation of parallelism} (cf.~Sections~\ref{subsec:scattered_memory_accesses} and~\ref{subsec:maximisation_of_parallelism}) could be addressed by providing a library of optimised functions and data structures for operations such as inter-agent communication or neighbour search (e.g.,~\cite{Coakley-2016,laville2013mcmas}).

A certain degree of \textbf{abstraction from hardware specifics} (cf.~Section~\ref{subsec:abstraction_from_hardware_specifics}) is achieved by the automated profiling and hardware mapping of our proposed workflow. Since each stage is executed on a single accelerator, facilities for unified memory access across all devices are not required. Instead, all agent data is updated locally on the accelerator and transferred automatically according to the schedule determined in the optimisation process.

Overall, the envisioned workflow is intended to rely on existing tools and techniques to allow researchers to exploit the hardware at their disposal with reasonable performance gains, while avoiding the need for costly and time-consuming manual optimisation steps as much as possible.

\section{Conclusions}
\label{sec:conclusions}

We presented a survey of the literature on agent-based simulation using hardware accelerators.
We categorized existing approaches according to the key challenges of hardware assignment, minimisation of data transfer overheads, scattered memory accesses, maximisation of parallelism, and the abstraction from hardware specifics.
Our survey provides modellers with an overview of techniques to execute a certain class of models on the available hardware.
Methodology researchers are given a summary of the existing work, pointing out research gaps where further exploration is required.
Our main observations are two-fold: first, most of the literature in the past years has focused on GPUs.
We expect a significant amount of work exploring agent-based simulations on FPGAs to appear in the near future.
Second, while a vast amount of work has proposed techniques that allow for efficient execution of agent-based simulations, only a small number of techniques has found their way into a unified framework.
Thus, the burden of developing a simulation that is executable in a heterogeneous environment is carried by the modeller.
Aiming to reduce the need for expertise in the programming for accelerators, we sketched our vision of a framework to perform an automated hardware mapping and performance optimisation based on building blocks from the literature.

\section*{Acknowledgement}
This work was financially supported by the Singapore National Research Foundation under its Campus for Research Excellence And Technological Enterprise (CREATE) programme.

\bibliographystyle{plain}
\bibliography{sim-opt-survey}

\end{document}